\documentclass[final,3p,times,twocolumn,authoryear]{elsarticle}

\usepackage{amssymb}
\usepackage{amsmath}
\usepackage{subfig}
\usepackage{enumitem}
\usepackage{xcolor}
\usepackage{lineno}
\usepackage{hyperref}
\usepackage{booktabs}
\usepackage{multirow}
\usepackage{physics}
\usepackage{lineno}
\usepackage{graphicx}
\usepackage{natbib}

\journal{Planetary and Space Science}

\begin{document}

\begin{frontmatter}


\title{On the formation of satellites in dense solid-particle disks}

\author[a,b]{G. Madeira}
\ead{madeira@on.br}
\author[c,b]{L. Esteves}
\author[a]{T.F.L.L. Pinheiro}
\author[c]{P.V.S. Soares}
\author[c]{N.S. Santos}
\author[d]{B. Morgado}

\affiliation[a]{organization={Observatório Nacional/MCTI},
            city={Rio de Janeiro, RJ},
            postcode={20921-400}, 
            country={Brazil}}
 
\affiliation[b]{organization={Université Paris Cité, Institut de Physique du Globe de Paris, CNRS},
            city={Paris},
            postcode={F-75005}, 
            country={France}}

\affiliation[c]{organization={Orbital Dynamics and Planetology Group (GDOP), São Paulo State University-UNESP},
            city={Guaratinguetá, SP},
            postcode={12516-410}, 
            country={Brazil}}

\affiliation[d]{organization={Universidade Federal do Rio de Janeiro - Observatório do Valongo},
            city={Rio de Janeiro, RJ},
            postcode={20.080-090}, 
            country={Brazil}}

\begin{abstract}
Single massive satellites are of great observational interest, as they can produce prominent and potentially detectable signatures. For terrestrial planets and super-Earths, giant impacts in the late stages of formation may generate dense self-gravitating disks -- favourable environments for the formation of such satellites. Motivated by this, we explore satellite formation in dense solid-particle disks through three-dimensional N-body simulations, focusing on the effects of disk mass and the surface density exponent ($\beta$). Our results reveal significant variability in the masses and configurations of satellites formed under identical disk parameters, highlighting the stochastic nature of the process. Higher disk masses and flatter surface density profiles favour the formation of more massive satellites. Disks with masses above 0.03 planetary masses typically yield a single dominant satellite, while those between 0.003 and 0.03 tend to form two-satellite systems. On average, the mass of the largest satellite scales linearly with the initial disk mass, in agreement with analytical predictions. We estimate that a disk with a minimal mass of 0.03 planetary masses around a 1.6 Earth-mass planet orbiting a Sun-like star could form an Earth–Moon-like system detectable by telescopes with a photometric precision of 10 parts per million -- a level achievable by the James Webb Space Telescope.

\end{abstract}



\end{frontmatter}


\section{Introduction} \label{sec_introduction}

In recent decades, space telescopes have detected thousands of exoplanets by monitoring stellar brightness variations during transits \citep{borucki2016kepler,gardner2023james}. Although exomoons have not yet been detected due to technological limitations \citep{teachey2024detecting}, their existence seems certain, given the recurrence of moons in our Solar System. Beyond the scientific curiosity, the discovery of exomoons is of great interest to astrobiology, as these objects may harbour liquid water and potentially support the development of life \citep{heller2013exomoon,heller2014formation}.

In our Solar System, the most populous satellite systems orbit the giant planets, where a first generation of regular satellites likely formed within gaseous circumplanetary disks during the solar nebula phase \citep[e.g.][]{canup2006common,canup2009origin,moraes2018growth,szulagyi2018situ,madeira2021building,anderson2021formation}. Subsequent generations may have arisen from the destruction, stripping, or disruption of these primordial satellites \citep{leinhardt2012tidal,crida2012formation, morbidellietal12,treffenstadt2015formation}. Similar processes are expected to take place around extrasolar gas giants planets, as circumplanetary disks have been detected around planets of this class \citep{christiaens2019evidence,benisty2021circumplanetary,sun2024observational}.

The terrestrial planets of our Solar System lack the mass to host circumplanetary disks \citep{szulagyi2017circumplanetary,dong2017mass}, and their satellites are generally attributed to giant impacts, which were common during the late stages of Solar System formation \citep{chambers01,raymondetal06,morishima2010planetesimals,morbidelli2012building}. The energy from such impacts typically vaporizes significant fraction of the impactor mass, creating a gas that expands and condenses into droplets \citep{hyodo2017impact}. These droplets coalesce into a debris disk, from which satellites form.

The Mars' satellites, for instance, are envisioned to be the final outcome of a disk-forming impact between Mars and a relatively small impactor with a mass of less than $\sim$3\% that of the planet \citep[e.g.][]{citron2015formation,rosenblatt2016accretion,hesselbrock2017ongoing,madeira2023exploring}. The Earth-Moon system, in turn, is proposed to have formed from a high-energy impact between Earth and a similar-mass body, which melted the silicate mantle and generated a disk from which the Moon later coalesced \citep[e.g.][]{canup2012forming,lock2018origin,lock2020energy,charnoz2021tidal,fu2024earth,madeira2025hydrodynamical}.

Pluto’s satellites may also have originated from a giant impact. This hypothesis is explored in \citet{canup2005}, which show that such an impact on Pluto typically produces a disk much less massive than Charon. Consequently, Pluto’s minor satellites may have coalesced from the debris disk \citep{stern2006,canup2011}, while Charon is more likely a direct fragment resulting from the impact.

Giant impacts are a natural consequence of planet formation in exoplanetary systems hosting super-Earths as well \citep[e.g.][]{chambers2013late,ogiharaetal18,carreraetal19,bierstekerschlichting19,lambrechtsetal19,estevesetal20,estevesetal22,izidoro2019formation,izidoroetal22,burnmordasini24,shibataizidoro25}. These exoplanets begin forming during the solar nebula phase by accreting their mass from either kilometer-sized planetesimals or centimeter- and millimeter-sized particles formed in the disk~\citep{birnstiel24,burnmordasini24,shibataizidoro25}. Once these planets grow to approximately one Earth mass, tidal interactions with the gas disk become strong enough to induce migration, typically resulting in planets being stacked in short-period orbits (\(a \lesssim 0.6\)~au). This process leads to the formation of populous and compact planetary systems~\citep{izidoro2017breaking,izidoroetal22,lambrechtsetal19}. After the gas disk dissipates, many of these systems undergo a phase of instability during which giant impacts occur multiple times, possibly giving rise to proto-satellite disks, from which exomoons might coalesce \citep{barr2017formation,malamud2020collisional}.

Motivated by this, we investigate the formation of satellites from massive solid-particle disks through N-body simulations. Previous studies on this topic \citep[e.g.][]{kokubo2000evolution,hyodo2015formation,sasaki2018particle} often rely on a single or very few simulations for each disk model, which limits their ability to fully capture the stochastic nature of the formation process. Here, we address this limitation by performing a large number of simulations for each disk model. Our methods are described in Section~\ref{sec_methods}, and we explore how the initial surface density profile influences on the system in Section~\ref{sec_surface}. The effects of the disk mass are analysed in Section~\ref{sec_mass}. Finally, Section~\ref{sec_discussion} presents a discussion of the results.

\section{Methods} \label{sec_methods}

To perform the numerical simulations of self-gravitating accretion disks, we use the N-body code \texttt{Rebound} \citep{rein2012rebound,tamayo2020reboundx}. The system is normalized such that the gravitational constant $G$ is set to unity, with distances and masses expressed in units of the central body's radius ($R_{\rm C}$) and mass ($M_{\rm C}$). Time is scaled so that $2\pi T_{\rm C} = 1$, where $T_{\rm C}$ is the Keplerian period at $R_{\rm C}$. The time step is set to $5 \times 10^{-3}$, which was determined through trial and error to reliably capture the dynamic scales of the particle disk. This value is comparable to that used in previous studies \citep{kokubo2000evolution,sasaki2018particle}, where it corresponds to $2^{-9}$ of their time unit $2\pi T_{RL}$ ($T_{RL}$ being the orbital period at Earth's Roche limit).

The code is configured to operate with multi-threading, utilizing both collision and gravity trees based on \citet{barneshut86} tree algorithm. Although the gravity tree introduces a loss of precision, comparisons between simulations with and without this feature revealed very similar dynamics with significantly lower computational costs. The tree code allows us to run multiple simulations with disks containing up to $N_p \sim 10^5$ particles within a reasonable time frame. In the following subsections, we describe the disk model and the collision treatment. 

\subsection{Disk Model}

We consider a disk with an initial mass $M_{\text{disk}}$, composed of 50000 equal-mass particles. This number ensures an approximation to the fluid-like behaviour of a disk -- clearly resolving structures such as spiral arms and clumps \citep{karjalainen2007,hyodo2014collisional,sasaki2018particle} -- while allowing for a large number of simulations. The assumed particle density is the bulk density of the Moon ($3.3 \, \text{g/cm}^3$), while the central body's density is taken as the Earth's ($5.5 \, \text{g/cm}^3$).

Impact simulations show that the majority of the mass in a disk formed by a giant impact is typically confined within the Roche limit of the planet \citep{cameron1997origin,canup2004simulations,citron2015formation,rosenblatt2016accretion,hyodo2017impact}. Based on this, we assume that the semi-major axes $a$ of the particles in the disk extend from the planet's surface to the Roche limit (\(a_{\rm RL}\)). Here, we compute the Roche limit assuming a perfect fluid approximation \citep{chandrasekhar1967ellipsoidal}, which yields a value of \(2.9 \, R_{\rm C}\).

The disk surface density is:
\begin{equation}
    \Sigma(a) = \Sigma_0 \left( \frac{a}{R_{\rm c}} \right)^{-\beta}
\end{equation}
\noindent where the exponent $\beta$ controls how dense the inner region of the disk is compared to the outer region, and $\Sigma_0$ is given by:
\begin{equation}
\Sigma_0 = \frac{M_{\text{disk}}}{\pi(a_{\rm RL}^2-R_{\rm C}^2)}\left[\int_{R_{\rm C}}^{a_{\rm RL}}\left( \frac{a}{R_{\rm c}} \right)^{-\beta} da\right]^{-1}
\end{equation}

The eccentricity and inclination distributions of the disk particles follow \citet{kokubo2000evolution}, where the root mean square eccentricity and inclination are $\langle e^2 \rangle^{1/2} = 0.3$ and $\langle i^2 \rangle^{1/2} = 0.15$, respectively. These values are set to mimic the orbits of the post-condensation debris \citep{cameron1997origin,canup2012forming,hyodo2017impact}. However, we assign lower eccentricities to particles in the inner region to prevent them from starting on collision courses with the central body.

As will be discussed later, we conducted a set of 30 simulations for each combination of \(M_{\text{disk}}\) and \(\beta\) parameters explored. These parameters were chosen because they directly affect the disk’s initial angular momentum, which primarily governs the mass of the largest satellite formed \citep{ida1997lunar}, and can be readily compared with giant impact simulations. While a more realistic particle size distribution than equal-mass particles is also expected to influence the initial angular momentum, it would introduce additional parameters (minimum and maximum particle masses), requiring a more detailed analysis that is beyond the scope of this study. Furthermore, comparing the particle size distribution with those obtained from SPH giant impact simulations is not straightforward, as the values obtained in SPH simulations depend on resolution and on how clumps are identified by the code.

To ensure that simulations with the same values of \(M_{\text{disk}}\) and \(\beta\) share the same initial angular momentum, we keep the semi-major axis, eccentricity, and inclination of the particles identical across simulations, varying only their angular orbital elements. The angular elements are randomly and uniformly distributed within the range \(0^{\circ}\) to \(360^{\circ}\), with any overlapping particles being relocated. Ensuring that simulations with identical parameters begin with same angular momentum allows for a fair comparison of results, as the initial angular momentum constrains the evolution of the disk \citep{ida1997lunar,kokubo2000evolution,canup2004simulations}.

\subsection{Collision Treatment}

We assume a hybrid treatment for collisions in our simulations, where they can be elastic or accretive depending on the impact conditions. Elastic collisions are treated using the hard-sphere approach already present in \texttt{Rebound}, with a tangential restitution coefficient of $\epsilon_t = 1.0$ and a normal restitution coefficient of $\epsilon_n=0.1$ \citep{kokubo2000evolution,sasaki2018particle}. Fragmentation is not considered in our simulations.

A gravitationally bound particle pair forms if the pair's escape velocity exceeds the rebound velocity of the collision. However, near the planet, this condition is modified by tidal forces, which act to stretch the pair \citep{ohtsuki1993capture,canup1995accretion}. Based on this, we assume a two-factor condition for an impact between two particles to be considered accretive: for two colliding particles $i$ and $j$, (i) the pair's Jacobi energy must satisfy $E_{\text{Jacobi}} < 0$, and (ii) the sum of their physical radii must satisfy $r_i + r_j~<~0.7~r_H$, where $r_H$ is the mutual Hill radius of the particles. The factor $0.7~r_H$ adjusts the theoretical Hill radius to account for the actual lemon-shape of the particle's Roche lobe in the proximity of the planet \citep{ohtsuki1993capture,canup1995accretion,kokubo2000evolution}.

The Jacobi energy of two colliding particles with a Keplerian angular velocity $\Omega$ is given by \citep{kokubo2000evolution}:
\begin{equation}
    E_{\text{Jacobi}} = \frac{v^2}{2} - \frac{3}{2}\Omega^2 x^2 + \frac{1}{2}\Omega^2 z^2 - G \frac{(m_i + m_j)}{r} + \frac{9}{2} r_{H}^2 \Omega^2
\end{equation}
\noindent where $r^2 = x^2 + y^2 + z^2$ is the square of relative position coordinates of particles $i$ and $j$, and $v$ is the relative velocity between the particles.

All impacts with the central body are treated as accretive, and particles are removed from the system when they exceed the distance of 50~$R_{\rm C}$. Our assumption of merging particles has the benefit of accelerating the simulations by reducing the number of particles over time; however, it is important to note that this approach may overestimate satellite masses, as it neglects potential mass losses caused by tidal stripping.

\section{On the influence of the initial surface density of the disk} \label{sec_surface}

We begin our study by analysing the effect of the initial surface density on the system, varying the parameter $\beta$. The initial disk mass is set to \( M_{\text{disk}} = 0.03~M_{\rm C} \) and we assume $\beta$=0, 0.5, 0.75, 3, and 5, performing a set of 30 simulations for each disk exponent value. $\beta = 0$ corresponds to a uniform surface density, while $0.5 \leq \beta \leq 5$ are typical values expected for accretion disks. For oblique giant impacts, such as those thought to have occurred on Earth, Pluto, and Mars, typical values are $\beta \gtrsim 1$ \citep{canup2004simulations,cuk2012making,meier2014origin,citron2015formation,hyodo2017impact,kenyon2021}. The timespan of our simulations is $10^4~T_{\rm C}$.

Using the conservation laws of the system, \citet{ida1997lunar} analytically determine that the mass of the largest satellite formed from a massive disk is primarily governed by the disk's initial angular momentum, a result numerically confirmed by \citet{ida1997lunar,kokubo2000evolution}. However, the stochastic nature of gravitational and collisional interactions among debris particles introduces a degree of chaos into the system \citep{hyodo2015formation}. Consequently, all our simulations with a fixed \(\beta\) value exhibit distinct evolutionary paths and final configurations, despite starting with the same angular momentum. For \(\beta = 5.0\), for instance, the largest satellite mass varies by an order of magnitude, ranging from 0.007 \(M_{\rm C}\) to 0.07 \(M_{\rm C}\), illustrating the strong influence of the stochastic nature of particle interactions on the final outcome \citep[see][]{hyodo2015formation,sasaki2018particle}.

\subsection{Comparison between two systems with identical disk parameters}

\begin{figure*}[]
\centering
\includegraphics[width=1.\textwidth]{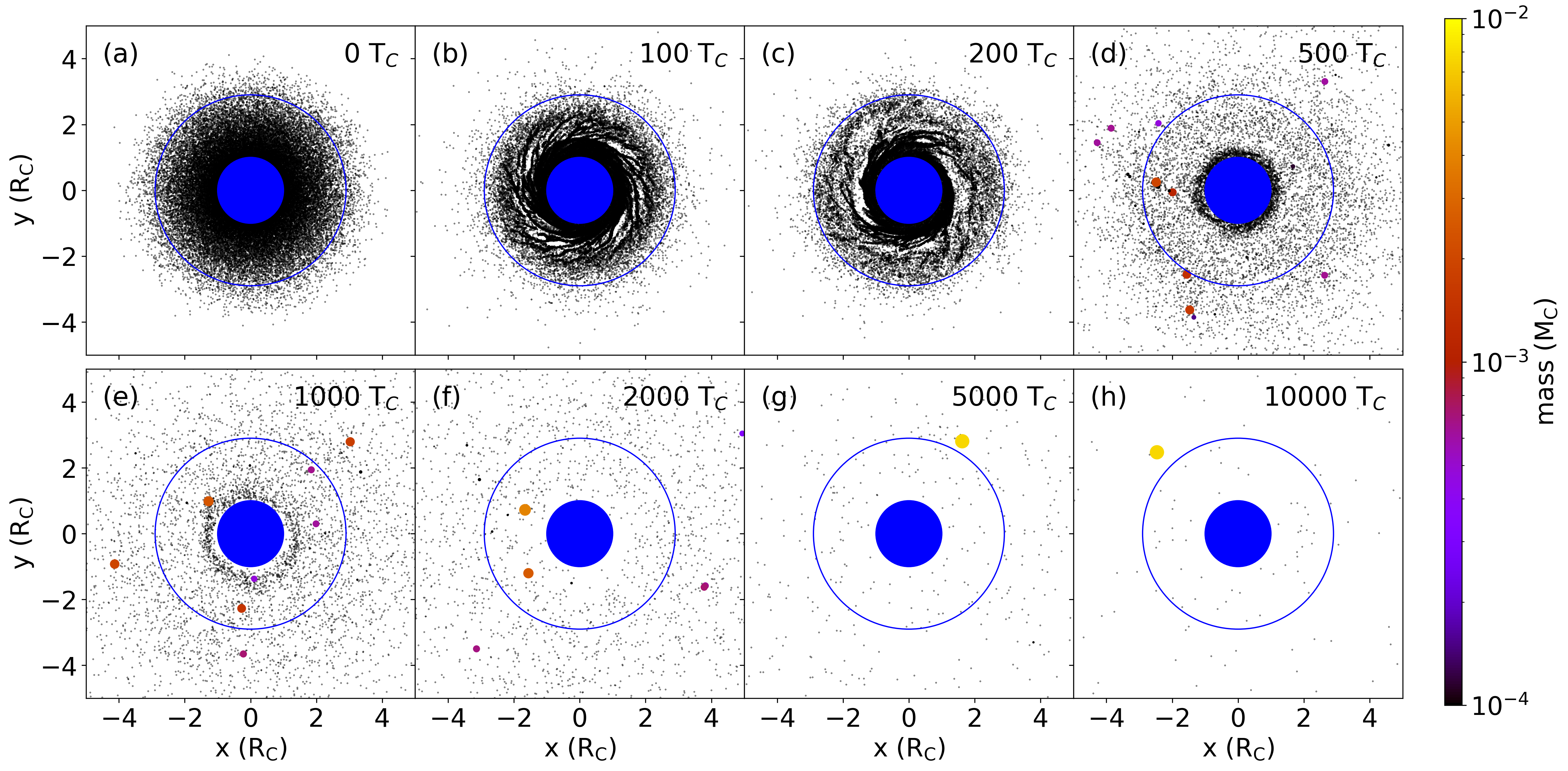}
\caption{Time series of a system initially with a disk of mass \( M_{\text{disk}} = 0.03~M_{\rm C} \) and an exponent value of \( \beta=0.75 \). The central body is shown in blue, with the solid blue line representing the planet's Roche limit. Disk particles are shown as black dots, while satellites are represented as coloured dots, with radii scaled to their physical sizes and colours coded by their masses.}
\label{fig:disks_11}
\end{figure*}

\begin{figure*}[]
\centering
\includegraphics[width=1.\textwidth]{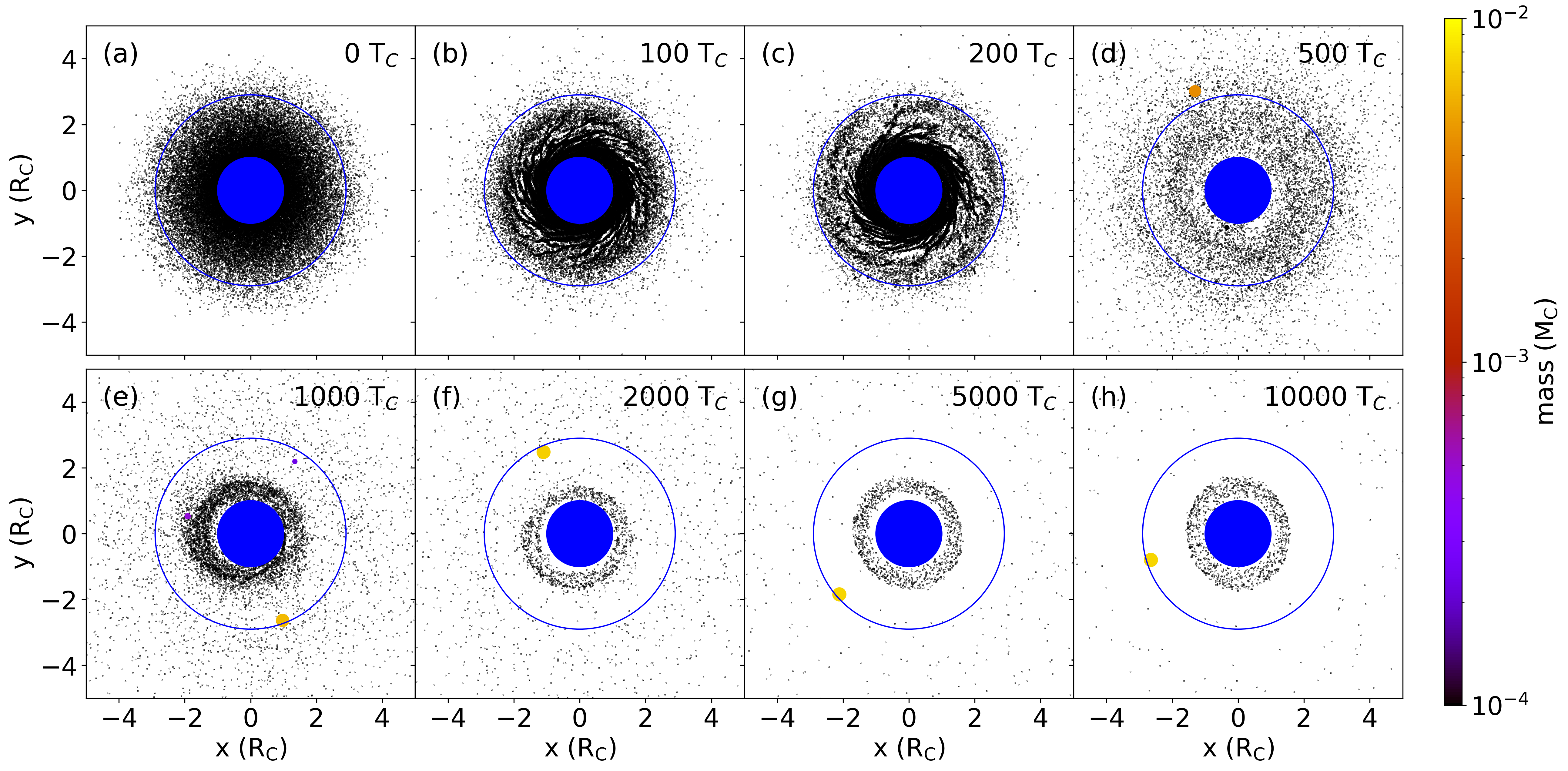}
\caption{Time series of a system with a disk of the same mass and surface density exponent as in Figure~\ref{fig:disks_11}, but with differently assigned values for the angular elements of the particles.}
\label{fig:disks_26}
\end{figure*}

\begin{figure}
\centering
\subfloat[]{\includegraphics[width=1.0\columnwidth]{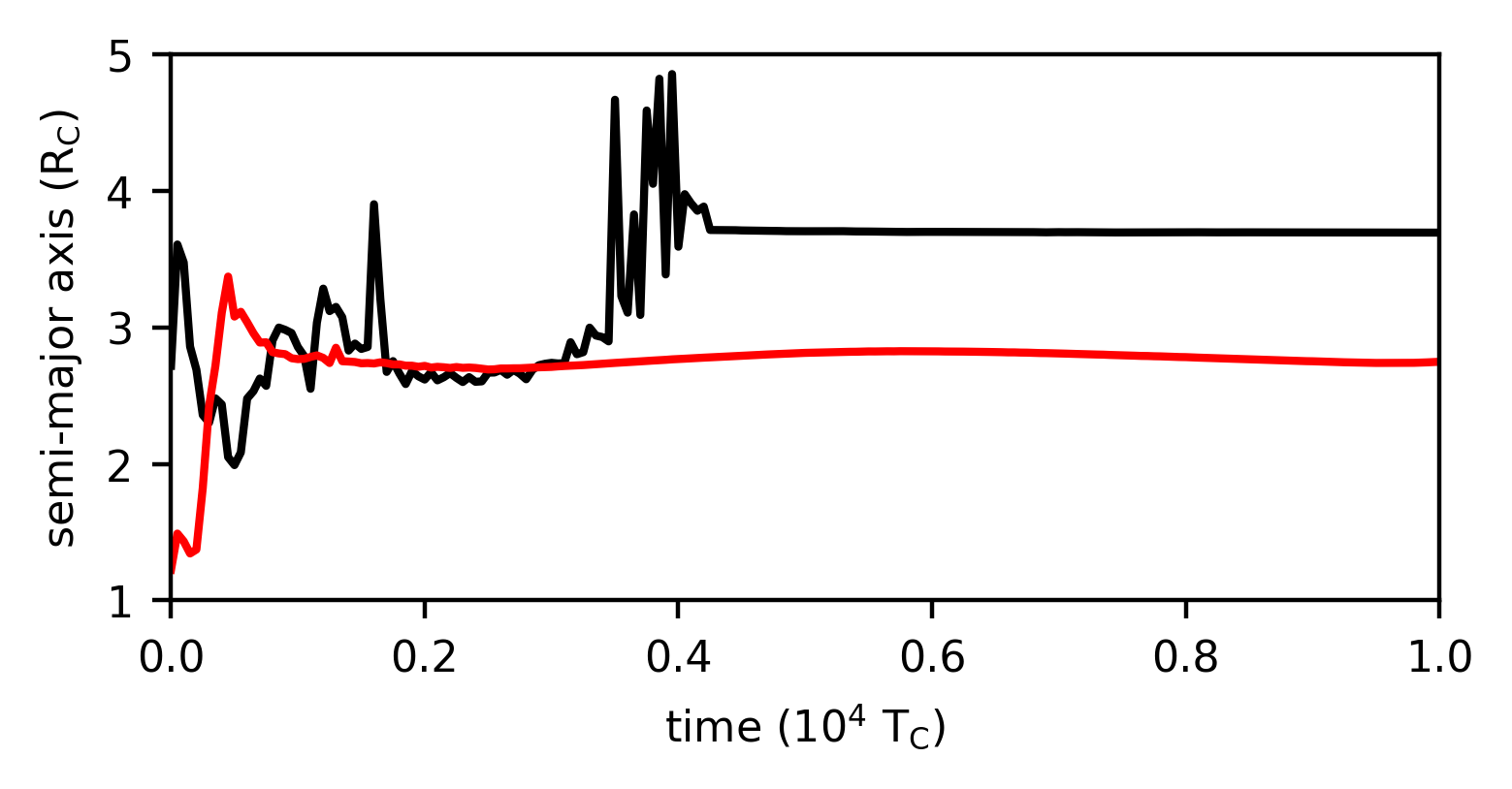}\label{fig:aeia}}\\
\subfloat[]{\includegraphics[width=1.0\columnwidth]{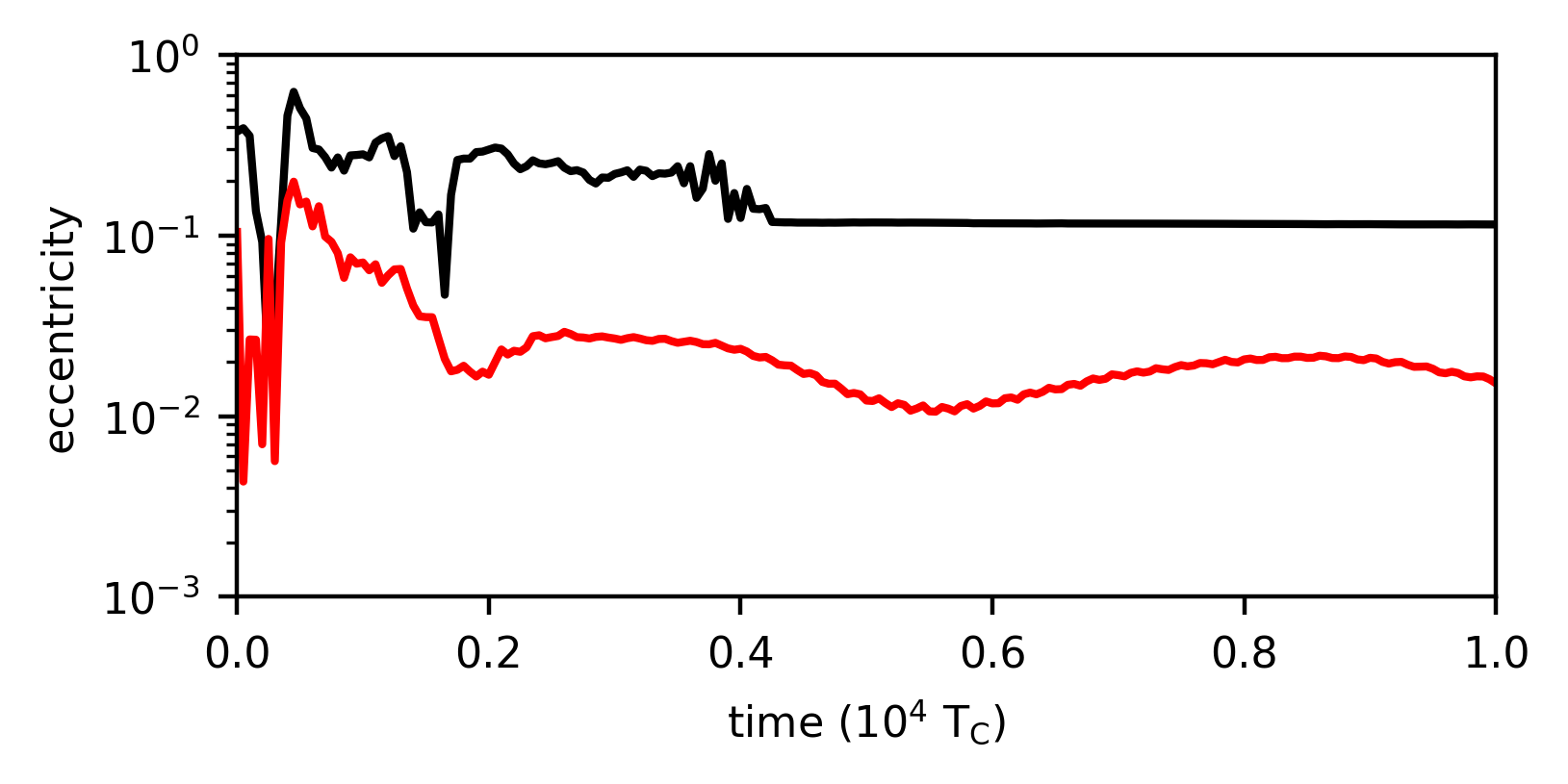}\label{fig:aeib}}\\
\subfloat[]{\includegraphics[width=1.0\columnwidth]{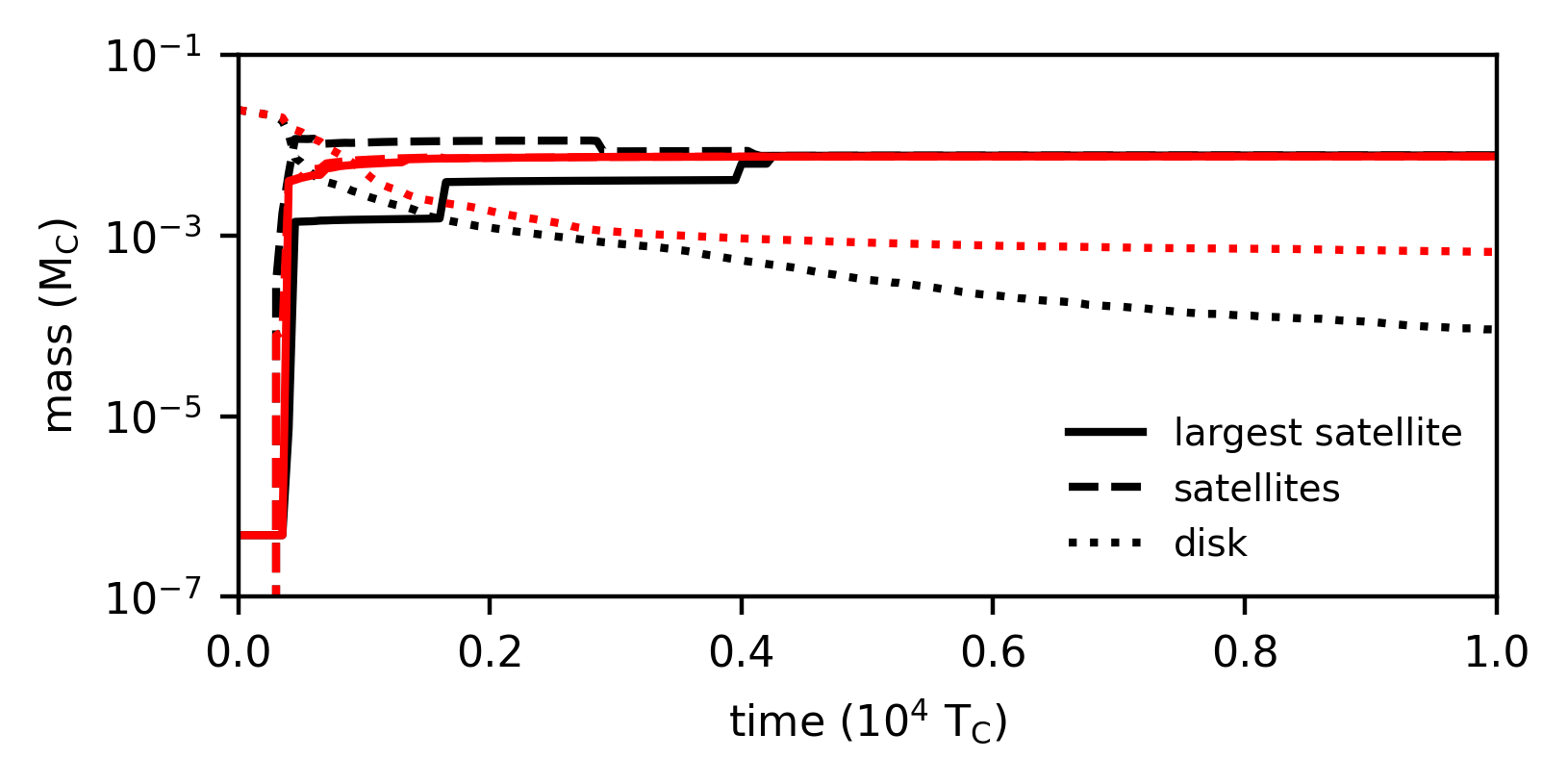}\label{fig:aeic}}\\
\subfloat[]{\includegraphics[width=1.0\columnwidth]{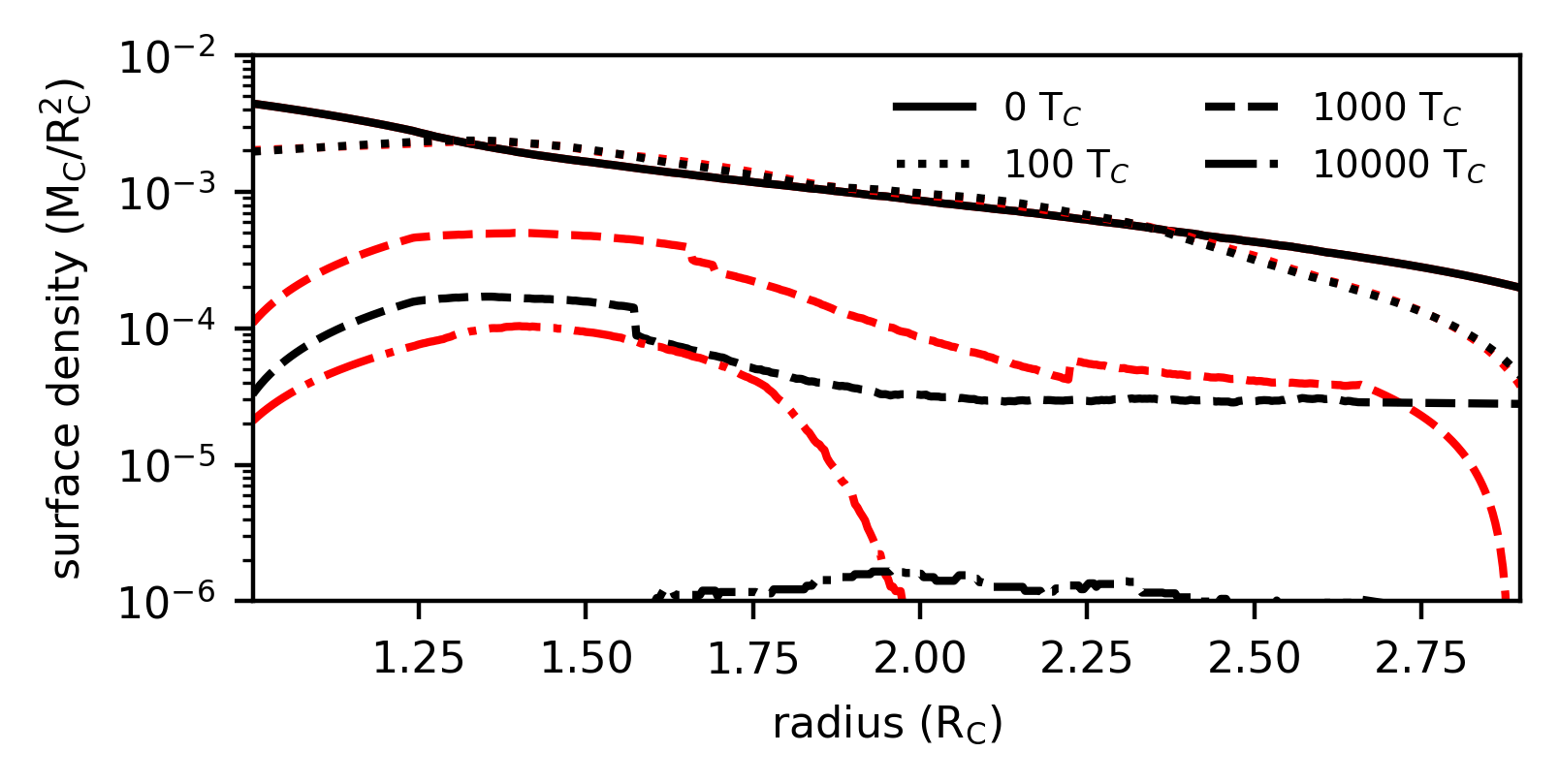}\label{fig:aeid}}
\caption{Temporal evolution of (a) semi-major axis, (b) eccentricity, and (c) mass of the largest satellite in the simulation. Panel (d) shows the disk surface density as a function of distance and time. The black and red lines stand for the simulations shown in Figures~\ref{fig:disks_11} and \ref{fig:disks_26}, respectively, both with $\beta=0.75$. In panel (c) we also display the disk mass and the total mass of satellites in the systems.}
\label{fig:aei}
\end{figure}

Figure~\ref{fig:disks_11} and Figure~\ref{fig:disks_26} present snapshots at different times for two simulations with $\beta = 0.75$, in which the final satellites have very similar masses despite following different formation pathways\footnote{Animations showing the evolution of these systems can be found at \url{gusmadeira.github.io/data}}. Temporal evolution of semi-major axis, eccentricity, and mass of the largest satellite formed in this pair of simulations is shown in Figure \ref{fig:aei}, which also shows the surface density of the disks as a function of distance and time.

The systems exhibit the well-known stages of formation in a gravitationally stable disk: (i) the disk contracts radially and vertically due to collisional damping (Figs.~\ref{fig:disks_11}a,~\ref{fig:disks_26}a); (ii) the disk surface density increases near the planet, causing the disk to become locally unstable at \(\sim 1.5 \, R_{\rm C}\); (iii) the first clumps coagulate near this position (Figs.~\ref{fig:disks_11}b,~\ref{fig:disks_26}b); (iv) the clumps are subsequently stripped by Keplerian shear, forming spiral arms (Figs.~\ref{fig:disks_11}b,c,~\ref{fig:disks_26}b,c); (v) the spiral arms transport mass to the outer regions of the disk due to angular momentum conservation, where the first aggregates form (Figs.~\ref{fig:disks_11}d,~\ref{fig:disks_26}d); and (vi) the aggregates combine, resulting in satellites (Figs.~\ref{fig:disks_11}d,e,~\ref{fig:disks_26}d,e)\footnote{We refer the reader to \citet{kokubo2000evolution,karjalainen2004gravitational,hyodo2015formation} for a detailed description of the early evolution of solid-particle disks.}. It is from stage (v) onwards that the two systems start to follow divergent evolutionary paths.

Here, we label as satellites objects with masses $\geq$5 particle masses, finding that the first surviving aggregates (stage v) coalesce around \(2.5 \, R_{\rm C}\) in both cases, within the Roche limit. This occurs at \(\sim 300 \, T_{\rm C}\) in both simulations, as evidenced by the sudden increase in the "satellites" curve in Figure~\ref{fig:aeic}. The coalescence of aggregates within the Roche limit is not surprising, given that the theoretical value relies on a perfect fluid approximation, while we are modelling solid particles evolving under dissipative collisions.

Dissipation is controlled by the normal restitution coefficient ($\epsilon_n = 0.1$), with lower values of $\epsilon_n$ corresponding to higher sticking probabilities. According to \citet{takeda2001angular}, the angular momentum transfer rates are largely unaffected by $\epsilon_n$, as long as $\epsilon_n \lesssim 0.6$\footnote{For $\epsilon_n \gtrsim 0.6$, dissipation becomes insufficient to reduce the random velocities of the particles, inhibiting the development of spiral structures and preventing aggregate growth \citep{takeda2001angular}.}. In this regime, systems show qualitatively similar dynamical evolution, although the satellite formation timescale varies.

In our simulations, aggregates form beyond $\sim 2.5 \, R_{\rm C}$. For higher values of $\epsilon_n$, we expect this formation region to shift further from the planet \citep{ohtsuki1993capture, karjalainen2004gravitational}, with satellite formation taking place over longer timescales. However, based on the findings of \citet{takeda2001angular}, we expect our statistical results (presented below) to remain robust across different values of $\epsilon_n$, as long as we remain in the dissipative regime defined by $\epsilon_n \lesssim 0.6$.

As already mentioned, the differences in the evolution of the two systems shown in Figures~\ref{fig:disks_11} and \ref{fig:disks_26} become striking after the coalescence of the first surviving aggregates, driven by their mutual gravitational interactions. During close encounters -- an important source of stochasticity in the system -- the aggregates tend to either scatter each other or collide. In the case of Figure~\ref{fig:disks_11}, a higher rate of scattering between aggregates is initially observed, and some aggregates survive independently. This can be noted in Figure~\ref{fig:aeic} by the difference between the dotted and solid black lines. Subsequently, these aggregates collide with each other, and the system ultimately evolves into a planet orbited by a single satellite and a small population of particles, leftover of the formation process.

In contrast, in the case of Figure~\ref{fig:disks_26}, the aggregates undergo collisions, merging into a single massive proto-satellite that subsequently grows by accreting disk particles. As the object grows, it clears its vicinity of particles and gravitationally dominates its surroundings. Once the satellite becomes sufficiently massive, it begins to shape the global structure of the disk through inner Lindblad resonances (ILRs) \citep{goldreich1980disk,meyer1987physics}. The satellite exerts a negative torque on the resonant segments of the disk, radially confining them -- which can be observed as step-like structures in Figure~\ref{fig:aeid}. In response, the satellite experiences a positive torque, causing its orbit to expand. 

By the end of the simulation (Fig.~\ref{fig:disks_26}h), the second system is composed of a single satellite and an eccentric ring confined within \(2 \, R_{\rm C}\), whose eccentricity mirrors that of the satellite \citep{winter2023stability}. As can be seen, the satellite ends up near the Roche limit, a configuration commonly observed in simulations that result in a confined ring. The final semi-major axis of the satellite is mainly affected by interactions with other satellites and by resonant torques exerted by the disk. In all simulations reported in this work, the semi-major axis of the largest satellite lies between \(2.5 \, R_{\rm C}\) and \(5.0 \, R_{\rm C}\). This upper limit is close to the location of the 2:1 ILR with the Roche limit (\(\sim 4.6 \, R_{\rm C}\)), which represents the maximum distance to which a satellite can migrate due to disk torques.

\subsection{Satellite mass as a function of $\beta$ parameter}

\begin{figure}
\centering
\subfloat[]{\includegraphics[width=1.0\columnwidth]{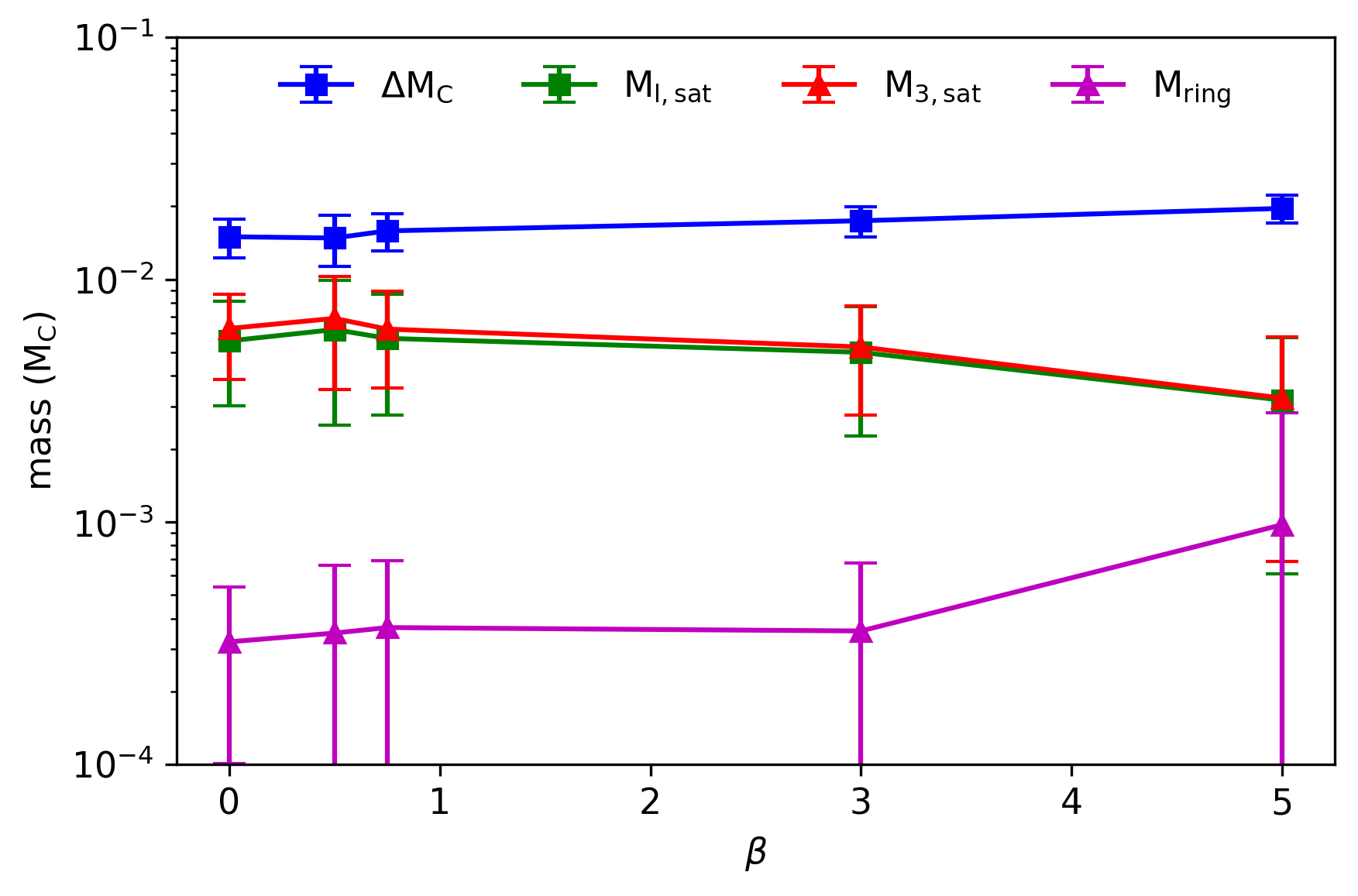}\label{fig:mass_decaya}}\\
\subfloat[]{\includegraphics[width=1.0\columnwidth]{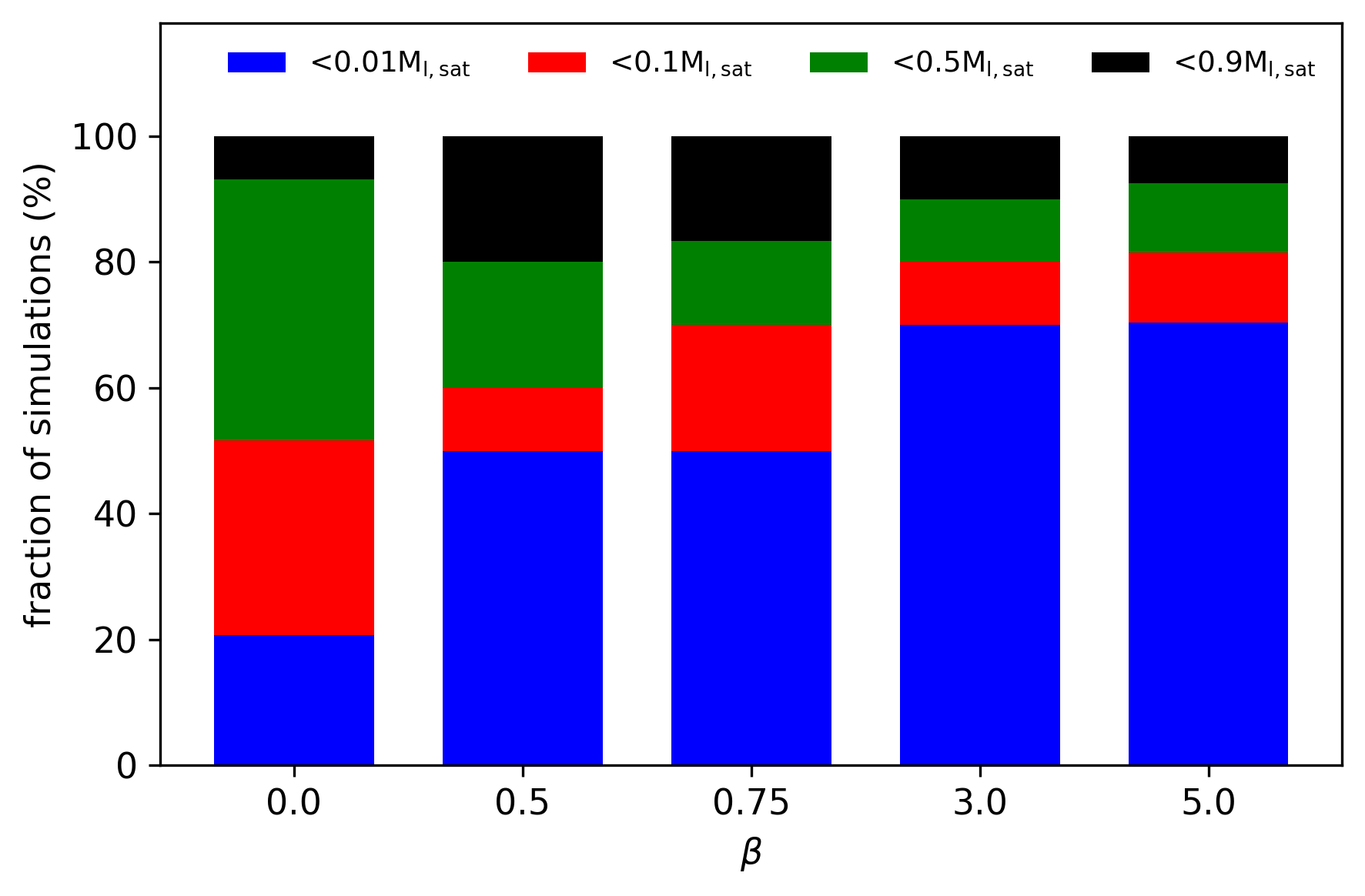}\label{fig:mass_decayb}}
\caption{(a) Mean values and standard deviations of the following quantities for different \(\beta\) values: mass fallen onto the planet (\(\Delta M_{\rm C}\), blue squares), mass of the largest satellite (\(M_{\rm l,sat}\), green squares), sum of the three largest satellites (\(M_{\rm 3,sat}\), red triangles), and disk mass (\(M_{\rm ring}\), pink triangles). (b) Fraction of simulations where the second largest satellite mass (\(M_{\rm 2l,sat}\)) is lower than 1\% (blue), 10\% (red), 50\% (green), and 90\% (black) of the largest satellite's mass.}
\label{fig:mass_decay}
\end{figure}

Figure~\ref{fig:mass_decay} shows the results of all our simulations for different \(\beta\) values. Fig.~\ref{fig:mass_decaya} presents the mean value and standard deviation of the final masses in the systems, while Fig.~\ref{fig:mass_decayb} shows the fraction of simulations where the mass ratio of the two largest satellites is less than 0.01, 0.1, 0.5, and 0.9. 

The stochastic nature of satellite formation is evident when analysing the standard deviation of the largest satellite's mass (\(M_{\rm l,sat}\)), which is of the same order of magnitude as the mean. For \(\beta = 0.75\), for instance, most simulations yield a largest satellite mass between \(3\times 10^{-3} \, M_{\rm C}\) and \(9\times 10^{-3} \, M_{\rm C}\), a variation of up to a factor of three. 

To confirm that satellite masses indeed span a broad range -- rather than clustering around discrete outcomes -- we examined the cumulative distribution functions of all simulation sets (including those in Section~\ref{sec_mass}) and compared them to different theoretical distributions (normal, log-normal, exponential, gamma) using the Kolmogorov-Smirnov test \citep{Chakravarti1967,estevesetal20,estevesetal22}. Among the tested models, the normal distribution provided the best fit, indicating that satellite masses are approximately Gaussian-distributed. This supports the use of the standard deviation in Figure~\ref{fig:mass_decay} as a meaningful measure of variability in satellite formation outcomes, reinforcing the conclusion that satellite masses span a broad range.

Despite the stochastic nature of the results, some patterns emerge. Disks with $\beta\geq2$ have a higher initial concentration of material near the planet. As the material spreads viscously \citep{salmon2010long}, more mass falls onto the planet (higher \(\Delta M_{\rm C}\)), reducing the mass available in the disk and, consequently, producing satellites with lower masses (\(M_{\rm l,sat}\) and \(M_{\rm 3,sat}\)). However, we point out that despite this decrease in satellite mass with \(\beta\), the mean value of \(M_{\rm l,sat}\) remains on the order of \(10^{-3} \, M_{\rm C}\).

The number of satellites formed is also observed to be influenced by the \(\beta\) parameter. Systems with $\beta<2$ generally lead to the formation of multiple satellites: the higher surface density in the outer regions of the disk allows several aggregates to coagulate and collide, reaching higher masses. In some cases, close encounters between aggregates scatter them into outer orbits, where they continue to grow by accretion. This process ultimately results in some systems with two or more satellites of comparable mass, as evidenced by the red, green, and black bars in Figure~\ref{fig:mass_decayb}. Nonetheless, unlike the results presented in the upcoming Section~\ref{sec_mass}, we do not observe any ranking of satellite mass with semi-major axis in systems with more than one satellite.

For $\beta\geq2$, the system normally ends up with only one satellite. This can be observed in the blue bars and the difference between the \(\Delta M_{\rm C}\) and \(M_{\rm 3,sat}\) curves in Figure~\ref{fig:mass_decay}. Due to the lower surface density in the outer disk, only one or a few aggregates form, which accrete the material available in their vicinity, clearing the region. The object then grows by accreting disk material that spreads outward. This process persists until the satellite becomes massive enough to confine or scatter the disk, halting further growth. Systems with $\beta\geq2$ more frequently exhibit a confined disk at the end of the simulations, which explains the increase in \(M_{\rm ring}\) with this parameter. In the next section, we examine the effect of the initial disk mass.

\section{Initial disk mass and satellite formation pathways} \label{sec_mass}


To investigate the impact of disk mass on the system, we conducted 30 simulations for each \(M_{\rm disk} = 0.003, 0.005, 0.008, 0.01, 0.03, 0.05, 0.08\), and \(0.1~M_{\rm C}\). For the disk surface density exponent, we follow \citet{sasaki2018particle} and adopt \(\beta = 3\). The timescale of satellite formation is directly proportional to the timescale of the viscous spreading of the disk \citep{crida2012formation,hyodo2015formation}, which, in turn, depends on the disk mass. Therefore, while in the previous section a timespan of \(10^4~T_{\rm C}\) was sufficient for all systems to reach equilibrium, we expect the formation timescales in this section to vary significantly across different sets of simulations. Seen this, we evolve all our simulations until \(M_{\rm ring} = 0.05 \, M_{\rm disk}\). This stopping criterion was determined through test simulations, which showed that satellites no longer exhibit significant mass growth when this ring mass is reached.

\begin{figure}
\centering
\subfloat[]{\includegraphics[width=1.0\columnwidth]{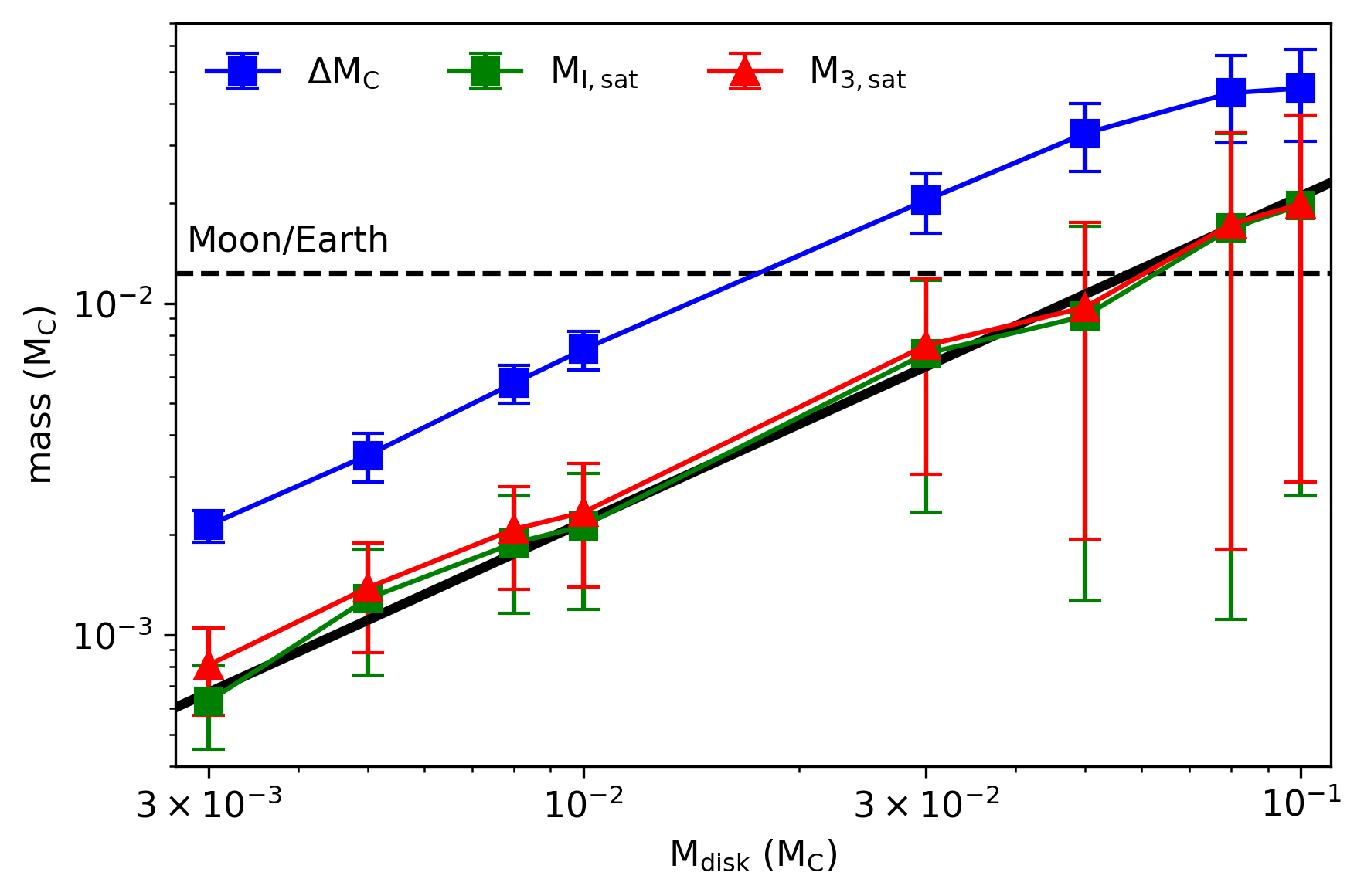}\label{fig:mass_massa}}\\
\subfloat[]{\includegraphics[width=1.0\columnwidth]{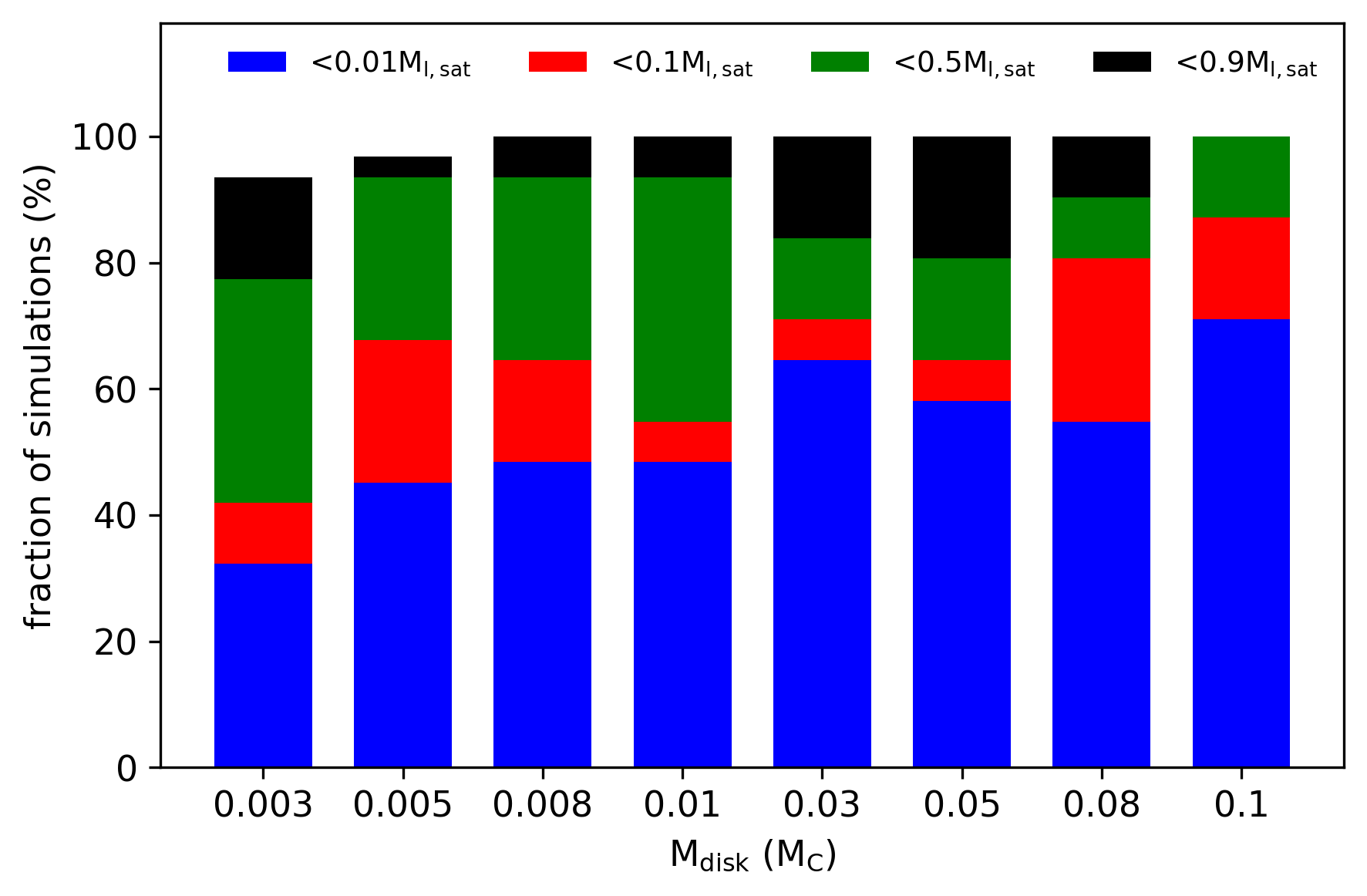}\label{fig:mass_massb}}\\
\caption{(a) Mean values and standard deviations of \(\Delta M_{\rm C}\), \(M_{\rm l,sat}\), and \(M_{\rm 3,sat}\); (b) fraction of simulations where the second largest satellite mass is less than different fractions of \(M_{\rm l,sat}\). In panel (a), the solid line represents our fit for the largest satellite mass as a function of the initial disk mass, while the dashed horizontal line gives the Moon-to-Earth mass ratio.}
\label{fig:mass_mass}
\end{figure}

Figure~\ref{fig:mass_massa} shows the mass that fall onto the planet, the mass of the largest satellite, and the combined mass of the three largest satellites across our different sets of simulations. Figure~\ref{fig:mass_massb}, on the other hand, presents the fractions of simulations where the mass ratio between the two largest satellites (\(M_{\rm 2l,sat}/M_{\rm l,sat}\)) is less than 0.01, 0.1, 0.5, and 0.9.

When the disk contains more mass, the mass flux toward the planet and the Roche limit increases. As a result, both the satellite mass and the mass accreted onto the planet increase with \(M_{\rm disk}\). 

Simulations with more massive disks also exhibit a greater variety of system outcomes, as seen in the increasing relative standard deviations of \(\Delta M_{\rm C}\), \(M_{\rm l,sat}\), and \(M_{\rm 3,sat}\) with \(M_{\rm disk}\). This is primarily driven by the gravitational perturbation and the resulting chaotic motion of disk particles, triggered by the first coalesced proto-satellite. The more massive the disk, the more massive the initial proto-satellite, which enhances perturbations and chaotic dynamics in the surrounding material \citep[see e.g.][]{Chirikov1979,henon1997,winter1997}.

Additionally, our adoption of more massive particles in higher-mass disk simulations further contributes to increase the stochastic behaviour of the system. More massive disks also lead to satellite formation on shorter timescales. For \(M_{\rm disk} \sim 10^{-3}~M_{\rm C}\), the largest satellite forms on timescales of \(10^4-10^5~T_{\rm C}\), while for \(M_{\rm disk} \sim 10^{-2}~M_{\rm C}\), the timescale decreases to \(10^3-10^4~T_{\rm C}\).

The initial disk mass also influences the number of satellites formed \citep{crida2012formation,hyodo2015formation}. In a massive disk, aggregates tend to undergo runaway growth, accreting particles within their Hill region. This process further expands the Hill region, enabling the aggregate to capture even more particles. With sufficient mass available in the disk, this process continues until the satellite becomes massive enough to scatter disk particles toward the planet, thereby preventing the formation of additional massive satellites. Consequently, massive disks tend to form single-satellite systems, like those shown in Figures~\ref{fig:disks_11} and \ref{fig:disks_26}.

\begin{figure*}[]
\centering
\includegraphics[width=1.\textwidth]{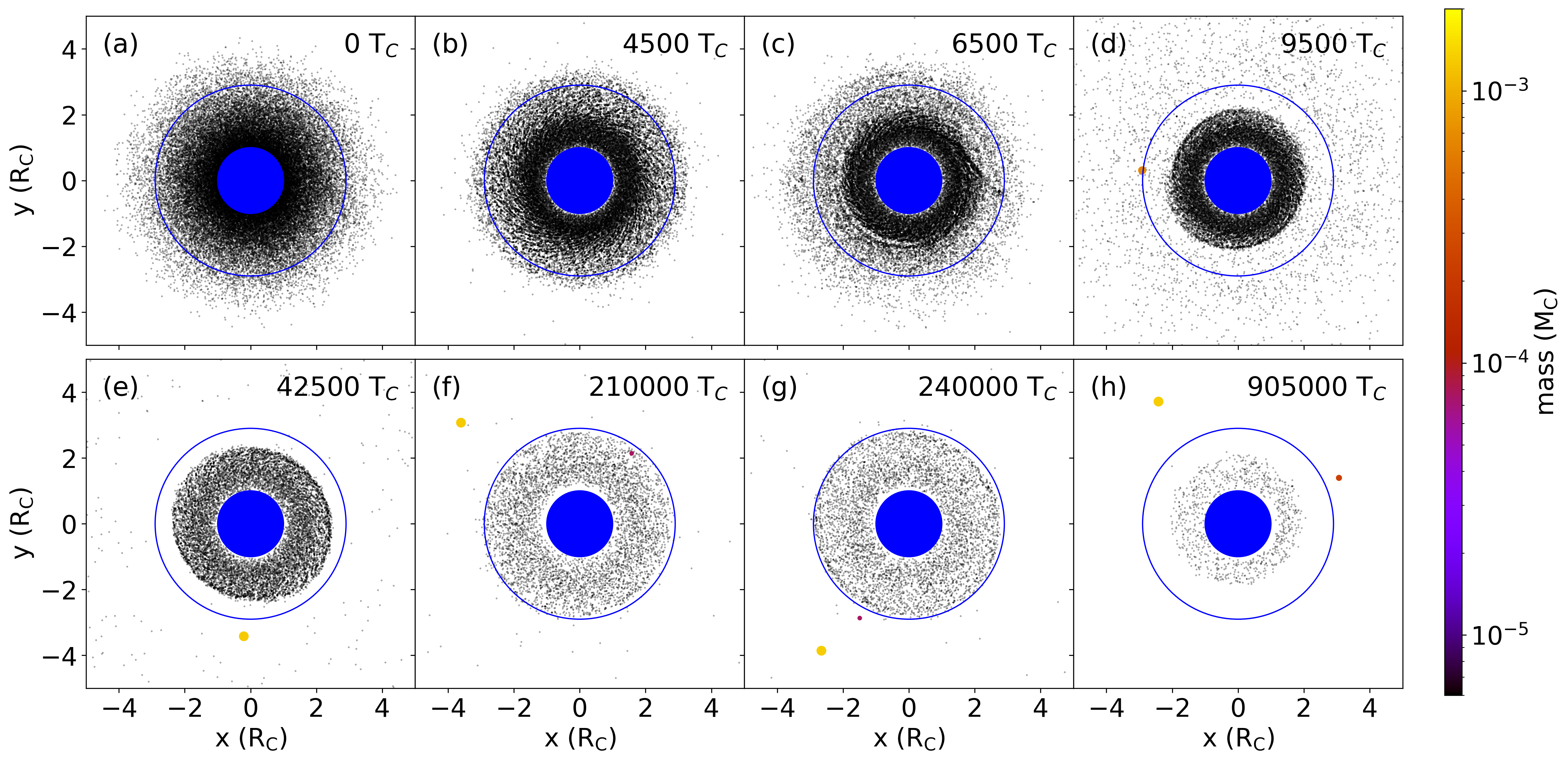}
\caption{Snapshots of a system initially with a low-mass disk (\(M_{\rm disk} = 0.005~M_{\rm C}\)). The different bodies in the system are represented by points, with sizes proportional to their physical radius and colours coded according to their masses. The central body is shown in blue, and the Roche limit is given by the blue circumference.}
\label{fig:disks_mass}
\end{figure*}

\begin{figure}
\centering
\subfloat[]{\includegraphics[width=1.0\columnwidth]{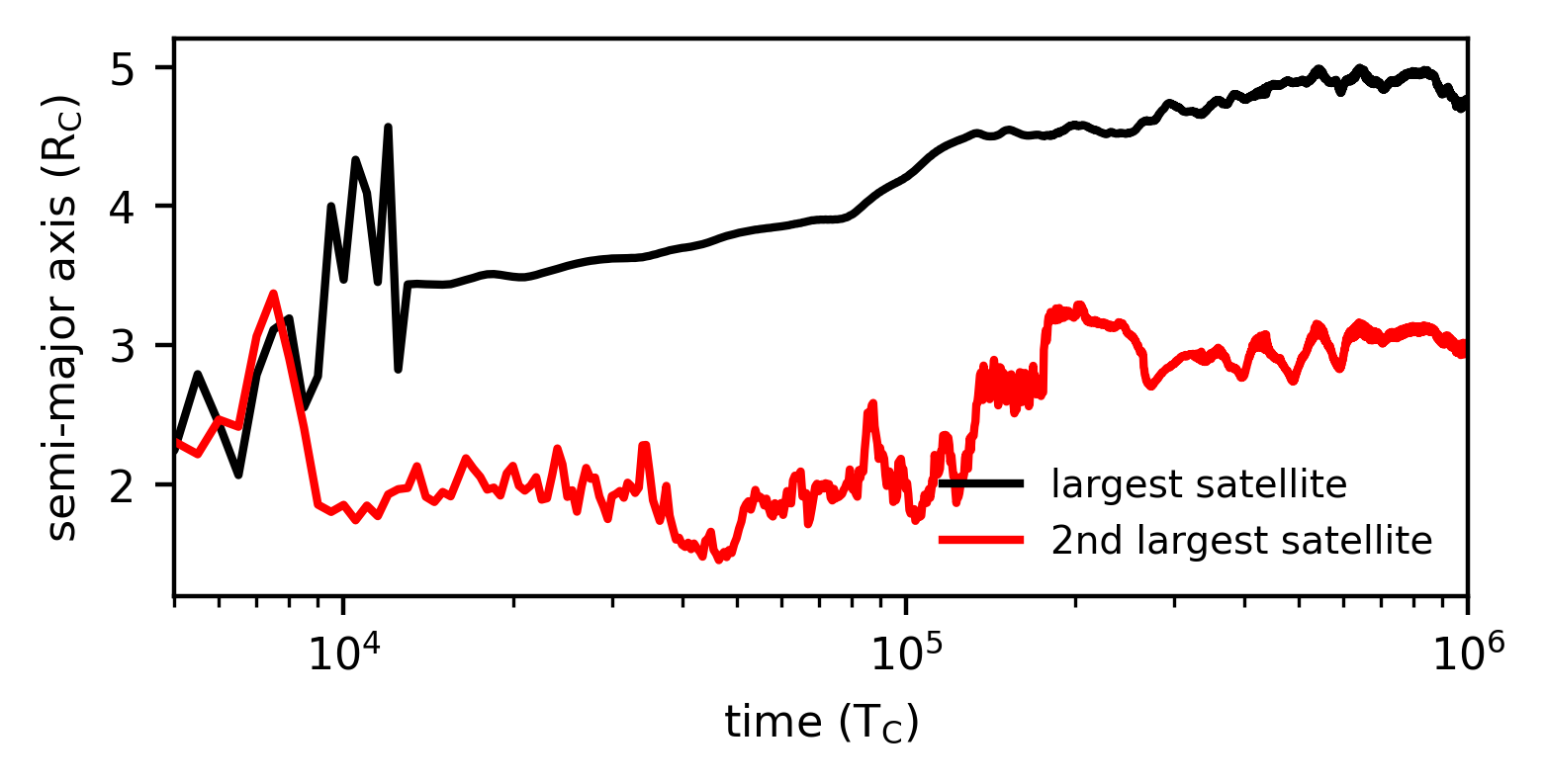}\label{fig:aei_massa}}\\
\subfloat[]{\includegraphics[width=1.0\columnwidth]{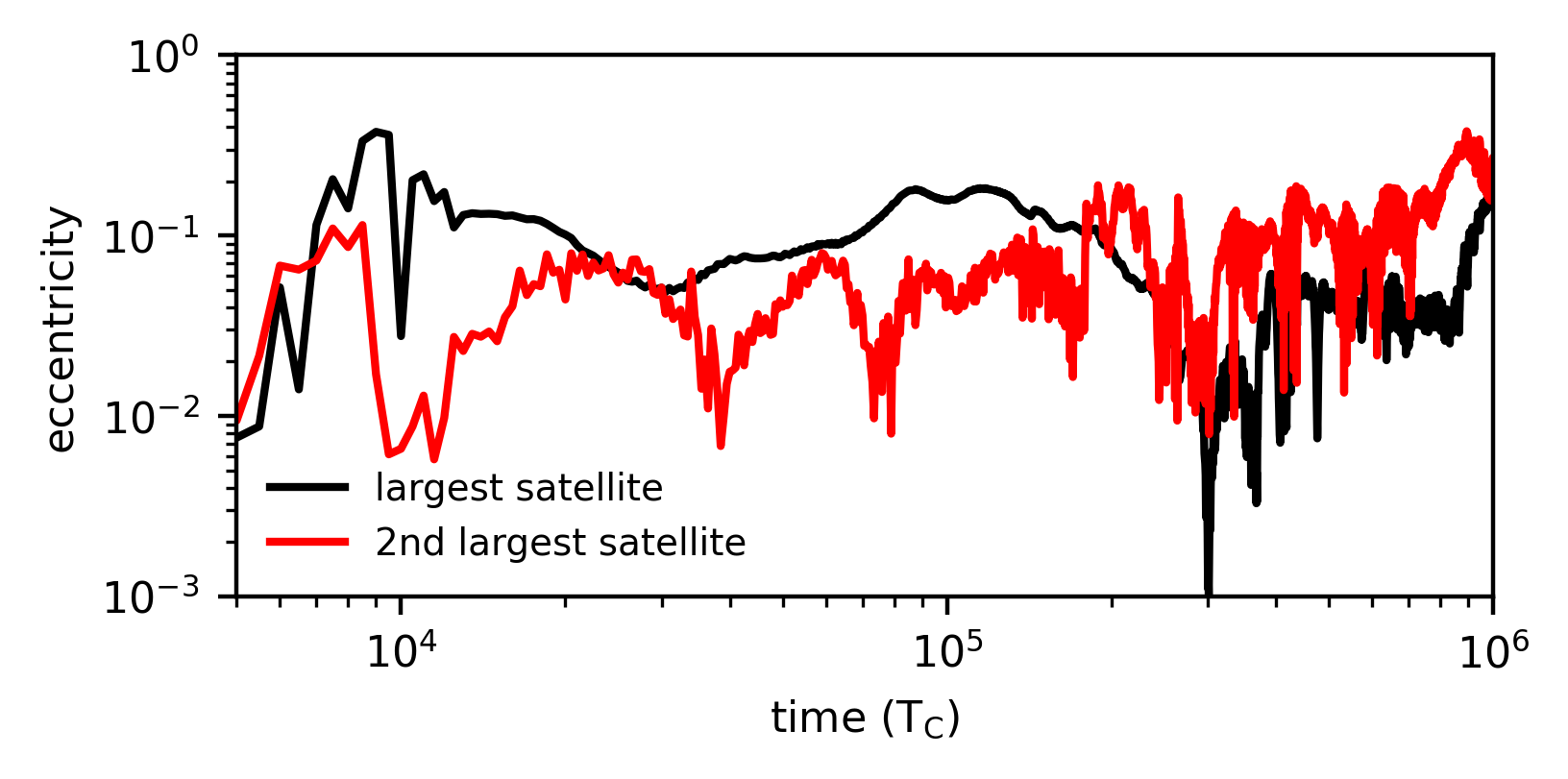}\label{fig:aei_massb}}\\
\subfloat[]{\includegraphics[width=1.0\columnwidth]{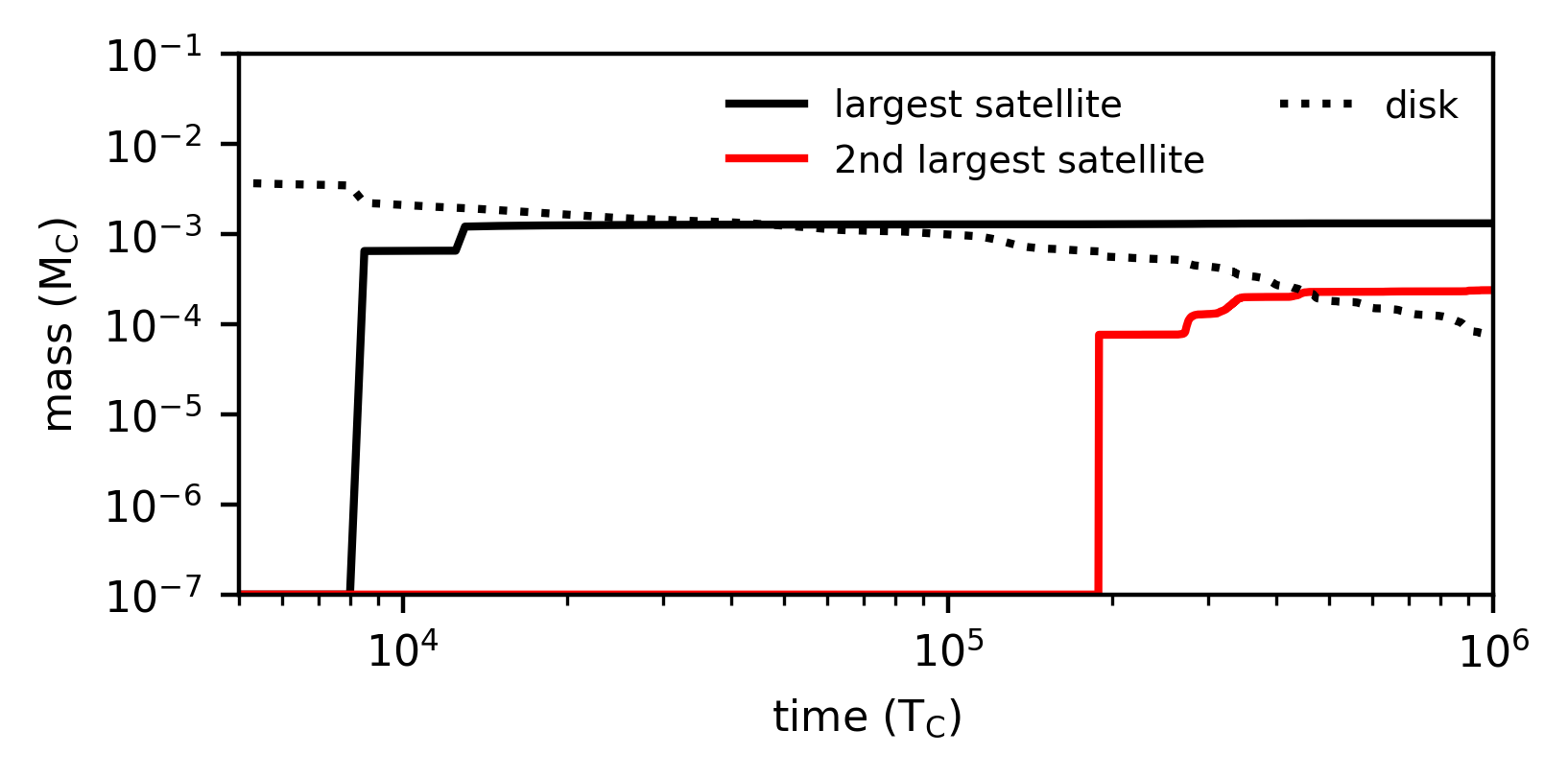}\label{fig:aei_massc}}\\
\subfloat[]{\includegraphics[width=1.0\columnwidth]{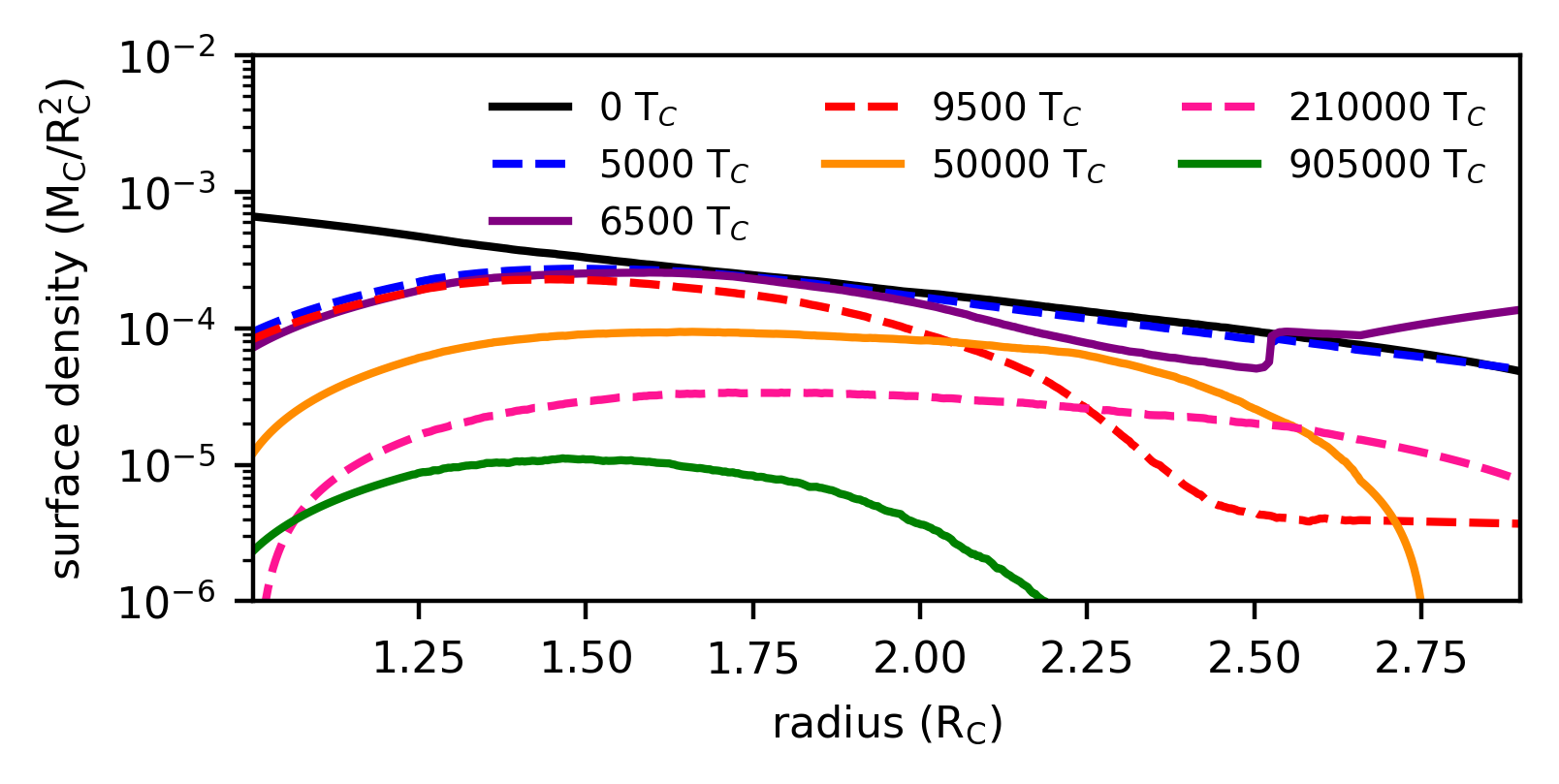}\label{fig:aei_massad}}\\
\caption{Evolution of (a) semi-major axis, (b) eccentricity, and (c) mass of satellites in the simulation shown in Figure~\ref{fig:disks_mass}, with the largest and second-largest satellites represented by black and red lines, respectively. In panel (c), the evolution of the disk mass is also shown as a dotted line. Panel (d) displays the disk surface density at selected times.}
\label{fig:aei_mass}
\end{figure}

For lower mass disks, the satellite does not become massive enough to clear out the disk, and we observe the formation of a second gravitationally dominant satellite. An example of a system with two satellites at the end is illustrated in Figure~\ref{fig:disks_mass}, where \(M_{\rm disk} = 0.005~M_{\rm C}\). Figure~\ref{fig:aei_mass} presents the temporal evolution of the semi-major axis, eccentricity, and mass of the satellites formed in Figure~\ref{fig:disks_mass}, along with the disk's surface density as a function of distance and time.

In this case, the first aggregate coalesces at \(6.5 \times 10^3~T_{\rm C}\), at the distance of \(2.5~R_{\rm C}\), as indicated by the step structure in the purple line in Figure~\ref{fig:aei_massad}. The system then evolves following the dynamics already discussed in \citet{crida2012formation}. The aggregate grows by accreting disk particles in its vicinity. Although the satellite does not become massive enough to clear the disk, it grows sufficiently to confine most of the disk mass within \(2.4~R_{\rm C}\) via 2:1 ILR -- as shown by the pink curve in Figure~\ref{fig:aei_massad} -- thereby preventing the immediate coalescence of new aggregates.

The satellite then accretes most of the unconfined particles and subsequently migrates outward due to disk torques. As the satellite's orbit expands, the outer edge of the disk also expands and eventually approaches the Roche limit, at which point a new aggregate coalesces, and the process repeats \citep{crida2012formation}. It is important to note that during the period when the disk remains confined within the Roche limit, spanning from \(10^4~T_{\rm C}\) to \(2 \times 10^5~T_{\rm C}\), disk particles are continuously accreted by the planet, leading to a reduction of the disk's mass over time. Consequently, the second satellite is always less massive than the first.

\citet{hyodo2015formation} identify that the transition between the predominant formation of a single massive satellite and the formation of two satellites occurs at \(M_{\rm disk} = 0.03~M_{\rm C}\), a result that is also confirmed by our study. For \(M_{\rm disk} < 0.03~M_{\rm C}\), the fraction of simulations where \(M_{\rm 2l,sat}/M_{\rm l,sat} = 0.1 - 1.0\) ranges from 50\% to 70\% (Figure~\ref{fig:mass_massb}), indicating that, in this disk mass interval, a second satellite typically forms with a mass approximately one order of magnitude smaller than that of the largest satellite. In contrast, in more than 50\% of the simulations with \(M_{\rm disk} \geq 0.03~M_{\rm C}\), \(M_{\rm 2l,sat}\) is at least 100 times smaller than \(M_{\rm l,sat}\). These systems can therefore be interpreted as producing only one gravitationally dominant satellite.

Intending to relate the typical mass of the largest satellite with the initial mass of the solid-particle disk, we applied different fits (power-law, log-log, exponential, linear, and polynomial) to our mean \(M_{\rm l,sat}\), using its standard deviation as a weight. We find that the curve that best fits our results is a simple power law (black solid line in Figure~\ref{fig:mass_massa}), given by:
\begin{equation}
M_{\rm l,sat}=(0.20\pm0.16)M_{\rm disk}^{0.98\pm0.16}. \label{eq_fit}
\end{equation}
This result indicates that the typical satellite mass is almost linearly proportional to the disk mass (\(M_{\rm l,sat} \approx (0.2\pm0.16)M_{\rm disk}\)), a relationship analytically predicted by \citet{ida1997lunar} (their Equation 2). However, the standard deviations are significant, reflecting the stochastic nature of the formation process. 

As a consequence of the system's stochastic nature, different initial disk masses can give rise to satellites with similar final masses. This is the case for the Moon, for which we find analogues in simulations with disk masses in the range \(0.03 \leq M_{\rm disk} \leq 0.1~M_{\rm C}\). This result is consistent with other studies on lunar formation from solid-particle disks \citep{kokubo2000evolution,hyodo2015formation,sasaki2018particle}, and also aligns with giant impact simulations, which typically yield disks with masses \(\sim 0.03~M_{\rm C}\) \citep{canup2012forming,cuk2012making,meier2024}.

In the case of Pluto, our fit (Equation~\ref{eq_fit}) indicates that forming a satellite as massive as Charon would require disks with masses \(\geq 0.3~M_{\rm C}\). This is significantly higher than the disk masses obtained in the simulations of \citet{canup2011} (\(\leq 0.05~M_{\rm C}\)), which strongly suggests that Charon did not form through accretion from a giant-impact-generated disk. In contrast, forming a satellite with the combined mass of Pluto’s minor satellites -- Styx, Nix, Kerberos, and Hydra -- requires disks with masses of \(\sim 10^{-4}~M_{\rm C}\), which is consistent with the disk masses obtained in the simulations of \citet{canup2011} that produces Charon as a direct fragment of the impact.

For the Mars system, we find that the direct formation of Phobos requires a disk with mass in the range \(\sim 10^{-7} - 10^{-6}~M_{\rm C}\). This range is close to that obtained by \citet{rosenblatt2012} using one-dimensional hydrodynamic simulations. It is also several orders of magnitude lower than the disk masses produced by impact simulations \citep{citron2015formation,rosenblatt2016accretion,hyodo2017impact}, which yield \(\sim 10^{-4} - 10^{-3}~M_{\rm C}\).

This discrepancy between the disk mass required for the direct formation of Phobos and that resulting from impact simulations has motivated the development of more sophisticated models for Phobos formation, such as the extended disk model \citep{rosenblatt2016accretion} and the recycling model \citep{hesselbrock2017ongoing}. It should be noted, nevertheless, that Equation~\ref{eq_fit} is only certainly suitable for disk masses within the range covered in this study. Therefore, the match with the results of \citet{rosenblatt2012} is likely coincidental.

\subsection{Satellite formation in low-mass disks}

In massive disks, such as those simulated here, satellites undergo runaway growth phase long enough to reach significant masses. However, as demonstrated by \citet{crida2012formation}, this process does not occur when the disk viscous spreading timescale becomes comparable to the planetary tidal effects, as is the case for low-mass disks. 

When the planetary tidal timescale is comparable to or shorter than the viscous spreading timescale, tidal dissipation accelerates the outward migration of satellites, halting their growth through particle accretion and promoting the sequential coalescence of new satellites at the Roche limit. Beyond the Roche limit, satellites grow through mutual collisions, ultimately forming a multi-satellite system with radial ranking of masses\footnote{This holds true only when the synchronous radius lies within the Roche limit, as both disk and tidal torques induce outward migration of the satellite. If the synchronous radius is outside the Roche limit, the satellite experiences opposing torques, resulting in more complex formation pathways where satellites can migrate both inward and outward, depending on the relative strength of each torque \citep{hesselbrock2019three}.} \citep[see also][]{madeira2023dynamical,madeira2024revisiting}.

To ensure the validity of our results, despite not including tidal effects in the simulations, we perform a simple comparison between the timespan for satellite formation in our simulations and the typical tidal migration timescale of the satellites (\(t_{\rm tide}\)). The tidal migration timescale is defined as \citep{goldreich1966q}:
\begin{equation}
t_{\rm tide} = \left(\frac{k_2}{Q}\right)^{-1} \frac{a^5}{3nM_{\rm l,sat}},
\end{equation}
where \(n\) is the Keplerian frequency of the satellite, and \(k_2/Q\) is the tidal parameter, which strongly depends on the internal properties of the planet, such as the structure of its core, mantle viscosity, and the presence of oceans or an atmosphere. In our calculations, we adopt \(k_2/Q \sim 10^{-3} - 10^{-2}\) as typical values for super-Earths and terrestrial planets \citep{tobie2005tidal,barr2018interior}, while the mass and orbital properties of the satellites are derived from our simulations.

The entire evolution of the system for cases with \(M_{\rm disk} \geq 0.03~M_{\rm C}\) occurs within a timespan of \(10^4~T_{\rm C}\). This duration is at least one order of magnitude shorter than the tidal migration timescale, which ranges from \(t_{\rm tide} \sim 10^5-10^6~T_{\rm C}\). Consequently, tides are expected to have a negligible impact on the formation process in these cases.

For \(M_{\rm disk} \sim 0.008-0.03~M_{\rm C}\), where the formation of a pair of satellites is predominant, the first and second satellites form on timescales of \(10^4~T_{\rm C}\) and \(10^5-10^6~T_{\rm C}\), respectively. The timespan for the second satellite formation is also one order of magnitude shorter than \(t_{\rm tide}\) (\(\sim 10^6-10^7~T_{\rm C}\)), suggesting that additional migration of the largest satellite due to tidal effects should not significantly interfere with the formation of the second satellite.

In contrast, for \(M_{\rm disk} \leq 0.005~M_{\rm C}\), the tidal timescale of the largest satellite (\(t_{\rm tide}\)) can become comparable to the timespan for the second satellite formation (\(10^6-10^7~T_{\rm C}\)), depending on the assumed \(k_2/Q\) value. Consequently, our results for low-mass disks should be interpreted with caution, as the formation process in these cases would be driven by planet's tides, which are not included in this study. We infer that terrestrial planets and super-Earths with disks of \(M_{\rm disk} < 0.003~M_{\rm C}\) will predominantly form satellites through the pathway described by \citet{crida2012formation}, likely resulting in systems with multiple satellites.

\section{Discussion} \label{sec_discussion}

In this work, we delve into the formation of satellites from dense solid-particle disks, a classic problem primarily explored in the context of the Moon formation. Here, however, we differentiate ourselves from previous studies \citep[e.g.][]{kokubo2000evolution,hyodo2015formation,sasaki2018particle} by performing a large number of simulations for each disk model. This approach allows us to capture the inherent stochastic nature of the satellite formation process, providing a more comprehensive understanding than prior studies, which often relied on fewer simulations. 

Some other publications also employ a larger number of simulations to study formation in a self-gravitating disk, such as \citet{wimarsson2024rapid}. However, these studies usually consider formation around small bodies with non-axisymmetric shape, which introduces additional chaos to the system \citep{madeira2022dynamics,ribeiro2023dynamics}. Here, for each set of values of disk mass \(M_{\rm disk}\) and surface density exponent \(\beta\) -- the key parameters of our model -- we conduct a total of 30 simulations, each with planet and a disk composed of 50,000 gravitationally interacting particles. 

For simplicity, we assume equal-mass particles in the disk, leaving the investigation of the effects of particle size distribution for future studies. We also do not account for flattening, density variations, or rotation of the central body. These factors are expected to influence the coagulation phase of satellite formation, depending on the characteristics of the bodies involved and the collision parameters. 

Our results reveal that the variability in satellite mass for the same set of parameters is substantial, increasing with disk mass. This is because more massive disks produce a more massive proto-satellite, which in turn induce chaotic motion over larger regions of the disk. As a result, the system exhibits a wider range of evolutionary outcomes, leading to larger variations in the final satellite mass.

For higher-mass disks (\(M_{\rm disk}\gtrsim0.05~M_{\rm C}\)), for example, the mass of the largest satellite varies by more than an order of magnitude from one simulation to another. Such variability is particularly relevant when modelling well-known and well-constrained systems, such as the Earth and Pluto systems. In such cases, even simulations of disks with identical initial mass and angular momentum may either reproduce or fail to match the observed satellite mass, due to stochastic effects. This highlights the need for statistical approaches in studying satellite formation from solid-particle disks, similar to those employed in studies of planet and satellite formation in gaseous disks \citep[e.g.][]{lambrechtsetal19,izidoro2019formation,madeira2021building,estevesetal22,izidoroetal22,burnmordasini24}.

Another factor that may influence the standard deviation of our results -- representing a caveat of our investigation -- is the limited number of particles in the disk. \citet{sasaki2018particle} demonstrate that very high-resolution simulations (N $= 10^7$) exhibit distinct wake-like structures absent at lower resolutions, which are associated with a global reduction in angular momentum transfer and slower satellite growth. Despite this, the final mass distribution among satellites, the disk, and material accreted by the central body in their simulations shows only minor variation with numerical resolution. We highlight, however, that these results are based on a very limited number of runs (up to two per resolution), which prevents firm conclusions due to the process’s inherent stochasticity. To limit the parameter space in this study, we do not explore the effects of particle number; however, its influence on satellite mass dispersion remains an open question that should be addressed in future work.

Despite the variety of outcomes, some patterns and typical formation pathways emerge from our simulations, and the disk mass is observed to have the most significant impact on system evolution. Similar to \citet{hyodo2015formation}, we identify two general formation pathways depending on the initial disk mass. Aggregates typically coalesce at \(\sim 2.5~R_{\rm C}\), near the Roche limit (\(2.9~R_{\rm C}\)), following a runaway growth phase. Nonetheless, it is the disk mass available near the aggregates that dictates the duration of this phase and the subsequent evolution of the system.

For high-mass disks, the runaway growth phase continues until the satellite scatters disk material or migrates far enough to halt accretion. This is precisely what is expected for the formation of the Moon, as our simulations yield final systems with a single satellite of mass comparable to the Moon's for initial disk masses in the range \(0.03\leq M_{\rm disk}\leq 0.1~M_{\rm C}\) and $\beta=3$. A similar conclusion applies to Charon, whose formation, according to our simulations, requires disk masses \(\geq 0.3~M_{\rm C}\). Such values are significantly higher than those inferred from impact simulations \citep{canup2011}, ruling out the possibility that Charon formed through accretion from a giant-impact-generated disk.

In lower-mass disks, runaway growth phase is shorter, and satellites do not scatter the disk. Nonetheless, within the mass range investigated here, they do become massive enough to confine the disk within \(2.5~R_{\rm C}\) via ILRs. This confinement temporarily halts satellite growth and the formation of new aggregates. Subsequently, the satellite's outward migration causes the disk to expand, leading to the coalescence of new aggregates and restarting the process.

Our simulations recover the findings of \citet{hyodo2015formation}, identifying \(M_{\rm disk}=0.03~M_{\rm C}\) as the threshold between these pathways. Disks more massive than this predominantly produce a single gravitationally dominant satellite, with additional satellites being at least two orders of magnitude less massive. For \(0.003 < M_{\rm disk} < 0.03~M_{\rm C}\), systems generally form two gravitationally dominant satellites, where the innermost satellite typically has 10–50\% of the mass of the largest one. A fit of the results demonstrates that, despite the variety of outcomes, the average mass of the largest satellite is nearly linearly proportional to the initial disk mass. This result aligns with the analytical predictions of \citet{ida1997lunar}, at least in the disk mass range explored in our study.

For even less massive disks, the processes described here are expected to repeat multiple times, culminating in systems with multiple low-mass satellites \citep{crida2012formation}. This is possibly what occurred in the case of Pluto's minor satellites. From Equation~\ref{eq_fit}, we infer that forming a satellite with the combined mass of Styx, Nix, Kerberos, and Hydra requires a disk with a mass of \(\sim 10^{-4}~M_{\rm C}\), consistent with the masses obtained in impact simulations where Charon emerges as a direct fragment \citep{canup2011}. This strongly supports such a scenario as the most plausible formation pathway for the Pluto system.

In such low-mass disks, viscous spreading is expected to occur over very long timescales (\(\gtrsim 10^6~T_{\rm C}\)), and planetary tides, an effect not included in our simulations, are expected to play a significant role in the formation process. For dissipative super-Earths and terrestrial planets, such as those with an oceanic layer, tidal effects might impact the formation process even for disks with masses \(\sim 10^{-3}~M_{\rm C}\). Planetary tides induce an outward migration of satellites, moving them farther from the disk. The implications of this effect for satellite formation are discussed in \citet{crida2012formation,hesselbrock2017ongoing,hesselbrock2019three,madeira2023exploring}, which assume an one-dimensional hydrodynamical approach for the disk.~A general evaluation of this effect in direct tridimensional N-body simulations should be further explored in future studies.

Our exploration of the surface density exponent demonstrates that this parameter also impacts the average masses of the satellites, as predicted by \citet{ida1997lunar}. Systems with single gravitationally dominant satellites are more common for higher \(\beta\) values ($\beta\gtrsim2$). This is because a higher initial concentration of material near the planet leads to a more substantial infall of material onto the planet, which reduces the disk mass and facilitates the scattering of the disk by the satellite. Whether this trend also holds for very low-mass disks remains to be verified.

Giant impacts are the most straightforward mechanism for driving satellite formation around extrasolar terrestrial planets and super-Earths \citep{barr2017formation,malamud2020collisional}. Impact simulations show that such events typically result in disks with $\beta\gtrsim1$ \citep{canup2004simulations,citron2015formation,cuk2012making,meier2014origin,hyodo2017impact,kenyon2021}. Consequently, our results indicate that exoplanetary systems with a single gravitationally dominant satellite are likely to be more common, provided the proto-satellite disk is sufficiently massive.

The formation of proto-satellite disks as a result of giant impacts has been investigated in cases involving planets with masses up to tens of Earth masses and impactor-to-total mass ratio in the range \(\sim 0.01-0.5\) \citep{barr2017formation,malamud2020collisional}. This mass ratio of colliding bodies aligns with those observed during the final stages of terrestrial and super-Earth formation \citep[e.g.][]{chambers2013late,estevesetal22}. The final state of the proto-satellite disks is shown to be highly sensitive to impact conditions, with disks forming masses ranging from \(\sim 10^{-4}-10^{-1}~M_{\rm C}\). The most massive disks (\(\sim 10^{-1}~M_{\rm C}\)) are the outcome of very oblique impacts (\(>50^{\circ}\)) with large impactor-to-system mass ratios (\(>0.1\)) and impact velocities near the system escape velocity.

The highly variable masses of proto-satellite disks, driven by their sensitivity to impact conditions, make it challenging to provide definitive constraints on realistic disk masses around terrestrial planets and super-Earths. Nonetheless, the results from \citet{barr2017formation,malamud2020collisional} lead us to envision that the formation of high-mass disks around these planets may not be uncommon. These disks, in turn, are likely to give rise to massive satellites, as highlighted in our study.

Moreover, the high-energy nature of these impacts introduces additional complexities, as the high temperatures associated with energetic giant impacts might lead to the formation of two-phase disks, depending on the impact conditions. In such cases, the evolution of the disk would not be driven solely by the gravitational and collisional interactions of its constituents but rather by the balance between heat dissipation and radiative cooling in the system \citep{ward2014evolution,charnoz2015evolution}.

\citet{Nakajima2022} find that, for impact parameters consistent with the proto-lunar-forming scenario \citep[e.g.][]{canup2012forming,cuk2012making,lock2018origin}, impacts onto planets larger than 1.6 Earth radii tend to produce vapour-rich disks. In such environments, km-sized moonlets that coalesce in the disk are expected to experience strong drag from the gaseous component, resulting in rapid inward migration and eventual infall onto the planet. This process likely inhibit the formation of massive satellites around planets in this size range, indicating that smaller planets are more favourable environments for their formation. While our simulations do not fully account for the dynamics of two-phase disks, the potential for massive satellite formation from vapour-poor disks remains of interest, particularly due to the observational signatures such systems may produce.

As a proof of concept, we estimate the minimum satellite mass required to produce a transit signal detectable by the James Webb Space Telescope (JWST), which achieves a photometric precision of approximately 10 ppm for bright stars \citep{bean2018}. Assuming a planet with the threshold radius of 1.6 Earth radii \citep{Nakajima2022} orbiting a Sun-like star, and considering a coplanar star–planet–satellite configuration, the maximum flux depth during a planetary transit can be expressed as:
\begin{equation}
    F = 1 - \frac{R_{\rm planet}^2}{R_{\rm star}^2} - \frac{R_{\rm satellite}^2}{R_{\rm star}^2} 
    \label{eq_flux}
\end{equation}
where $R_{\rm star}$, $R_{\rm planet}$, and $R_{\rm satellite}$ are the radius of the planet, star, and satellite, respectively.

Using this relation, we find that a satellite must have a mass exceeding $0.006~M_{\rm C}$ to be detectable by JWST. According to Figure~\ref{fig:mass_decay}a, such a satellite could form from a disk with a mass of around $0.03~M_{\rm C}$, which is comparable to estimates for the proto-lunar disk \citep{canup2012forming,cuk2012making,meier2024}. Given that the Earth–Moon system is an outlier among planet–satellite systems in the Solar System, this may help explain the current lack of exomoon detections. Nevertheless, it also suggests that the discovery of exomoons may soon be within reach, particularly with the development of telescopes offering greater photometric precision than JWST.

In addition to causing a depth in transit flux, massive satellites displace the centre of mass of the planet-satellite system, resulting in changes to the timing of transits and the duration of transit events \citep{sartoretti1999detection,szabo2006possibility,kipping2009transit,barr2017formation}. The search for these signatures is the main detection method employed by the Hunt for Exomoons with Kepler \citep{kipping2012hunt}, one of the leading initiatives dedicated to the search for moons outside the Solar System. Therefore, the possibility of proto-satellite disks forming as a natural outcome of terrestrial and super-Earth planet formation, and the subsequent formation of massive satellites, presents an intriguing observational opportunity as larger satellites produce more prominent and potentially detectable signatures.

\section*{Declaration of competing interest}
The authors declare that they have no known competing financial interests or personal relationships that could have appeared to influence the work reported in this paper.

\section*{Acknowledgements}
G. Madeira acknowledges support from the Institut de Physique du Globe de Paris and the European Research Council (grant 101001282, METAL). L. Esteves acknowledges funding from the Fundação de Amparo à Pesquisa do Estado de São Paulo (FAPESP) under grants 2023/09307-3 and 2021/00628-6. P.V.S. Soares, N.S. Santos and T.F.L.L. Pinheiro acknowledge support from the Coordenação de Aperfeiçoamento de Pessoal de Nível Superior - Brasil (CAPES) - Finance Code 001. We thank the anonymous reviewers for their thoughtful comments and valuable insights on this work. Part of the numerical simulations was performed on the S-CAPAD/DANTE platform at IPGP, France, while another part was carried out using the computational resources of the Center for Mathematical Sciences Applied to Industry (CeMEAI), funded by FAPESP (grant 2013/07375-0), and the computational resources of the Grupo de Dinâmica Orbital e Planetologia, funded by FAPESP (grant 2016/24561-0). This work was developed within the framework of the research project conducted by the Group of Origin and Dynamics of Satellites.

\section*{Data availability}
The data underlying this article will be shared on reasonable request to the corresponding author.

\bibliographystyle{plainnat}

\end{document}